\documentclass[manuscript]{aastex}
\usepackage{natbib}
\usepackage{lineno}
\usepackage[figuresright]{rotating}
\usepackage{lscape}

\shorttitle{Peculiar burst of 4U~1608--52}
\shortauthors{Chen et al.}

\begin{document}

\title{Insight-HXMT observations on thermonuclear X-ray bursts from 4U~1608--52 in the low/hard state:
the energy-dependant hard X-ray {\bf deficit} and cooling saturation of the corona}

\author{Yu-Peng Chen\textsuperscript{*}}
\email{chenyp@ihep.ac.cn}
\affil{Key Laboratory for Particle Astrophysics, Institute of High Energy Physics, Chinese Academy of Sciences, 19B Yuquan Road, Beijing 100049, China}

\author{Shu Zhang\textsuperscript{*}}
\email{szhang@ihep.ac.cn}
\affil{Key Laboratory for Particle Astrophysics, Institute of High Energy Physics, Chinese Academy of Sciences, 19B Yuquan Road, Beijing 100049, China}

\author{Long Ji\textsuperscript{*}}
\email{jilong@mail.sysu.edu.cn}
\affil{School of Physics and Astronomy, Sun Yat-Sen University, Zhuhai, 519082, China}

\author{Shuang-Nan Zhang}
%\email{zhangsn@ihep.ac.cn}
\affil{Key Laboratory for Particle Astrophysics, Institute of High Energy Physics, Chinese Academy of Sciences, 19B Yuquan Road, Beijing 100049, China}
\affil{University of Chinese Academy of Sciences, Chinese Academy of Sciences, Beijing 100049, China}

\author{Jing-Qiang Peng}
\affil{Key Laboratory for Particle Astrophysics, Institute of High Energy Physics, Chinese Academy of Sciences, 19B Yuquan Road, Beijing 100049, China}
\affil{University of Chinese Academy of Sciences, Chinese Academy of Sciences, Beijing 100049, China}

\author{Ling-Da Kong}
%\email{kongld@ihep.ac.cn}
\affil{Key Laboratory for Particle Astrophysics, Institute of High Energy Physics, Chinese Academy of Sciences, 19B Yuquan Road, Beijing 100049, China}
%\affil{University of Chinese Academy of Sciences, Chinese Academy of Sciences, Beijing 100049, China}
\affil{ Institut f\"{u}r Astronomie und Astrophysik, Kepler Center for Astro and Particle Physics, Eberhard Karls Universit\"{a}t, Sand 1, D-72076 T\"{u}bingen, Germany}

\author{Zhi Chang}
\affil{Key Laboratory for Particle Astrophysics, Institute of High Energy Physics, Chinese Academy of Sciences, 19B Yuquan Road, Beijing 100049, China}

\author{Qing-Cang Shui}
\affil{Key Laboratory for Particle Astrophysics, Institute of High Energy Physics, Chinese Academy of Sciences, 19B Yuquan Road, Beijing 100049, China}
\affil{University of Chinese Academy of Sciences, Chinese Academy of Sciences, Beijing 100049, China}

\author{Lian Tao}
\affil{Key Laboratory for Particle Astrophysics, Institute of High Energy Physics, Chinese Academy of Sciences, 19B Yuquan Road, Beijing 100049, China}

\author{Jin-Lu Qu}
\affil{Key Laboratory for Particle Astrophysics, Institute of High Energy Physics, Chinese Academy of Sciences, 19B Yuquan Road, Beijing 100049, China}
\affil{University of Chinese Academy of Sciences, Chinese Academy of Sciences, Beijing 100049, China}

\author{Ming-Yu Ge}
\affil{Key Laboratory for Particle Astrophysics, Institute of High Energy Physics, Chinese Academy of Sciences, 19B Yuquan Road, Beijing 100049, China}

\author{Jian Li}
\affil{CAS Key Laboratory for Research in Galaxies and Cosmology, Department of Astronomy, University of Science and Technology of China, Hefei 230026, China}
\affil{School of Astronomy and Space Science, University of Science and Technology of China, Hefei 230026, China}

\begin{abstract}

During thermonuclear bursts, it is suspected that
{\bf the cooling of the corona by the burst emission}
 may be the cause of hard X-ray {\bf deficits}. Although this {\bf deficit} has been observed in nine sources, it has not been observed {\bf from} 4U~1608--52, a nearby prolific burster. Therefore, the authenticity and universality of the hard X-ray {\bf deficit} may be in question. To investigate this suspicion, Insight-HXMT performed cadence observations
during the low/hard state of 4U~1608--52 in September 2022 and detected 10 thermonuclear X-ray bursts. Two of these bursts show a double-peaked structure in the soft X-ray band, which could be caused by the high temperature of the burst emission and a marginal photospheric radius expansion (PRE) around the burst peak time. This is indicated by their peak fluxes being up to the Eddington limit and having a large color factor at the peak of the bursts.
The hard X-ray deficit is significantly observed during bursts at $>$ 30 keV.
Furthermore, the fraction of this deficit shows saturation at 50\% for the first 8 bursts. This saturation may indicate that the corona is layered and only a part of the corona is cooled by the bursts. For example, the part close to the NS surface is cooled while the rest remains intact during bursts.
This result provides a clue to the geometry of the corona, e.g., a possible scenario is that the corona has two forms: a quasi-spheric corona between the NS and the disk, and a disk-corona on both surfaces of the disk.
\end{abstract}



%%%%%%%%%%%%%%%%%%%%%%%%%%%%%%%%%%%%%%%%%%%%%%%%%%

%%%%%%%%%%%%%%%%% BODY OF PAPER %%%%%%%%%%%%%%%%%%

\onecolumn
\section{Introduction}

{\bf
The thermonuclear bursts, also known as type-I X-ray bursts (hereinafter referred to as bursts),  occur in neutron star (NS) low mass X-ray binaries (LMXBs) due to the thermonuclear burning of accreted material onto the NS surface (see a review, see, e.g., \citealp{Galloway2020}).
These bursts appear as short-duration flashes superimposed on the persistent (i.e., non-burst) emission, characterized by a rapid rise and slow decay light curve, spectrum softening during the decay phase, and burst waiting time on the order of hours.
The most luminous burst can reach a peak flux up to the Eddington luminosity, potentially resulting in photospheric radius expansion (PRE).
As the burst emerges from the NS surface, its spectrum and lightcurve can be used to estimate NS parameters, such as mass, radius, spin and surface gravity.

The persistent emission originates from several regions, including the NS surface, the accretion disk, the corona and the boundary/spreading layer (BL/SL), where the accretion flow decelerates from the Keplerian period to the spin period of the NS.
The BL is believed to spread over a large radial extent in the disk midplane, while the SL has a narrower spread but extends over a considerable height from the equatorial plane toward higher stellar latitudes.
Criteria for distinguishing BLs and SLs have been established based on temporal \citep{Gilfanov2003} and spectral \citep{Grebenev2002,Suleimanov2006} properties.
%In theory, in a high accretion rate with the BL/SL between the accretion flow and the surface, the  energy deposited in the BL/SL should be greater than that in the accretion disk. For example,  for the fastest confirmed millisecond pulsar PSR~J1748--2446  (at 716 Hz), the energy released in the BL should  be as much as that in the accretion disk (e.g., \citealp{Done2007}).
However, identifying these regions is still challenging due to the spectral degeneracy and telescope constraints.
%The two types of X-ray emission mentioned above could mutually influence each other.
The interactions between thermonuclear bursts and accretion circumstances could provide insights into decomposing these emission regions.
For example, the burst emission could cool the corona via Compton scattering of soft burst photons with energetic electrons, leading a decline in the hard X-ray lightcurve during bursts (e.g., \citealp{maccarone2003, Chen2018,2020MNRAS.499.4479S}).

In literature, nine bursters with a hard X-ray {\bf deficit} have been observed (see Table \ref{tb_burster}).
The {\bf deficit} fraction of the hard X-ray emission is approximately 20\%--100\%. In general, it is energy dependent, having larger values in higher energy bands.
However, a hint of the {\bf deficit} saturation ($\sim$ 50\%) at 50--60 keV \citep{Chen1816} was detected from MAXI~J1816--195, which contradicts the theoretical prediction that a burst cooling the corona to a lower temperature should result in a larger deficit fraction in higher energy bands.
}

In addition, the hard X-ray {\bf deficit} has not been detected in 4U 1608-52, a nearby transient X-ray source with a distance of 2.9--5.4\,kpc  \citep{Galloway2020}, either by RXTE/PCA \citep{Ji2014b} or NuSTAR \citep{2016MNRAS.456.4256D}.
This source exhibits prolific outbursts and bursts, with activity occurring every several months and an inferred burst rate of $\sim$ two bursts per day.
%This has prompted an investigation into the authenticity and universality of the hard X-ray {\bf deficit} detected during bursts from 4U~1608--52, and cadence observations were conducted during the low/hard state by Insight-HXMT in 2022.
Thus, to test the authenticity and universality of the hard X-ray {\bf deficit}, we performed an extensive study for bursts occurring in the low/hard state of the outburst in 2022 using cadence Insight-HXMT observations.

\section{OBSERVATION AND DATA ANALYSIS}

\begin{figure}
\centering
\includegraphics[angle=0, scale=0.6]{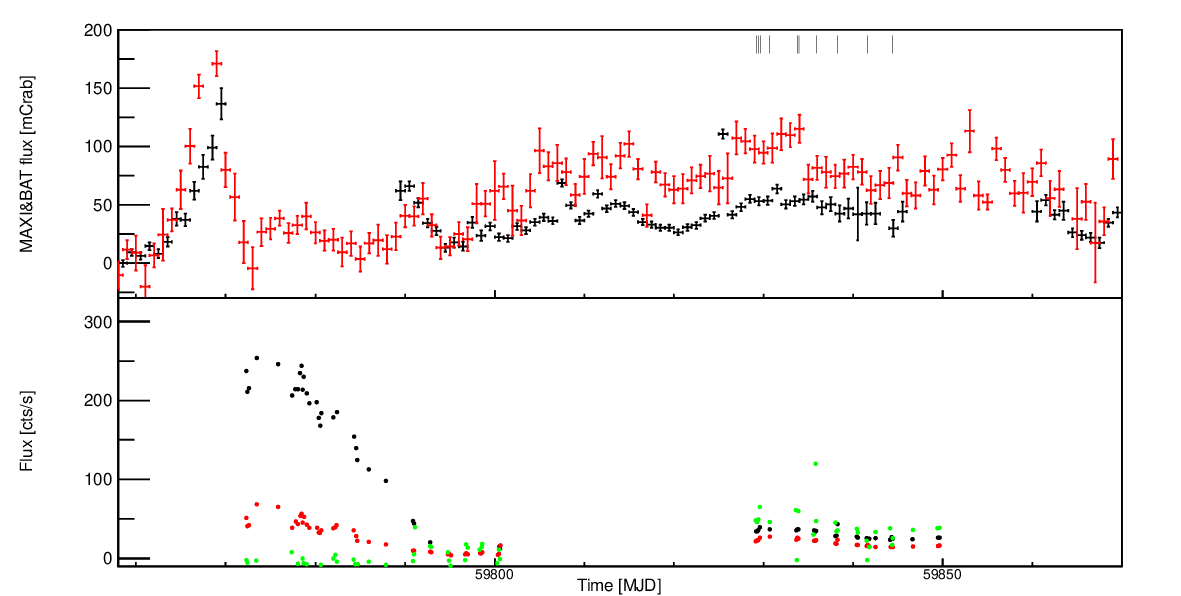}
  \caption{Top panel: daily light curves of 4U~1608--52 by MAXI (black, 2--20 keV) and Swfit/BAT (red, 15--50 keV) during the outburst in  2022. The  bursts are indicated by vertical lines. Bottom panel: light curves of 4U~1608--52 by LE (green), ME (blue), and HE (red) which are rebinned by one obsid ($\sim$ 10000 s). Please note the error bars of the light curves are smaller than the size of the symbols.
  }
\label{fig_outburst_lc}
\end{figure}

\begin{figure}
\centering
\includegraphics[angle=0, scale=0.6]{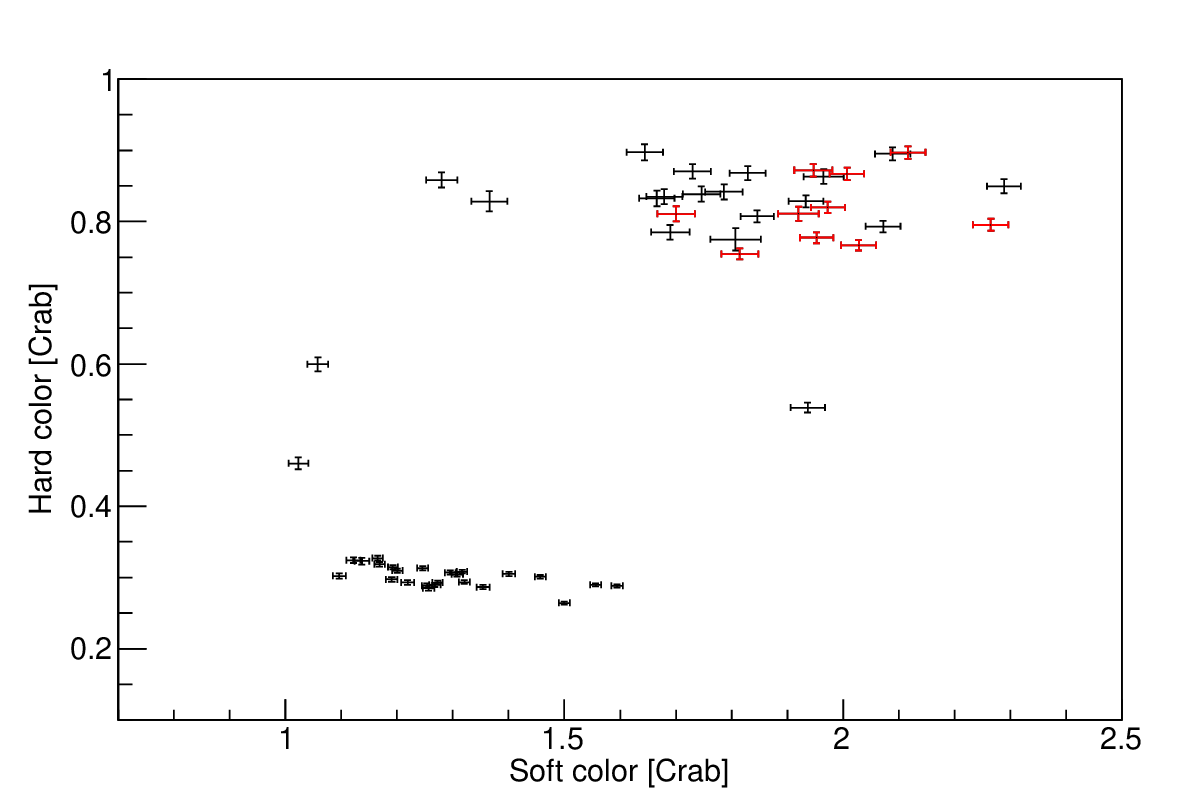}
  \caption{The color--color diagram (CCD) of this outburst. The soft color is defined as the ratio of count rate in 6--10 keV \& 1--6 keV of LE and the hard color is defined as the ratio of count rate in 10--15 keV \& 15--30 keV of ME. Both of them are normalized by the count rates of Crab at the same energy bands and at the same   period.  The outburst moves in a  clockwise direction: begins at the bottom right  and stops at the top right. {\bf The red data points  indicate the observations where the thermonuclear bursts occurred.}}
\label{fig_outburst_ccd}
\end{figure}

Insight/HXMT \citep{Zhang2019} performed 26 observations of 4U~1608--52 from 2022 September 7th to 27th with obsids P050422200101-20220907-01-01 to P050422201302-20220927-01-01, as shown in Figure \ref{fig_outburst_lc}.
The observational data from the Low Energy X-ray Telescope (LE), the Medium Energy X-ray Telescope (ME) and High Energy X-ray Telescope (HE), which are sensitive in 1--10 keV, 8--30 keV and 25--250 keV, respectively, are used for analysis.
These observations accumulated a total unfiltered exposure time of 293 ks.
The lightcurves and spectra of the three telescopes were extracted using three pipelines of HXMTSOFT v2.05: lepipeline, mepipeline and hepipeline.
To track the outburst evolution, the color-color diagram (CCD) was plotted (Figure \ref{fig_outburst_ccd}) based on the ratio of count rates in 6–10 keV and 1–6 keV (soft color) of LE and the ratio of count rates in 10--15 keV and 15--30 keV (hard color) of ME.

\begin{figure}
\centering
\includegraphics[angle=0, scale=0.2]{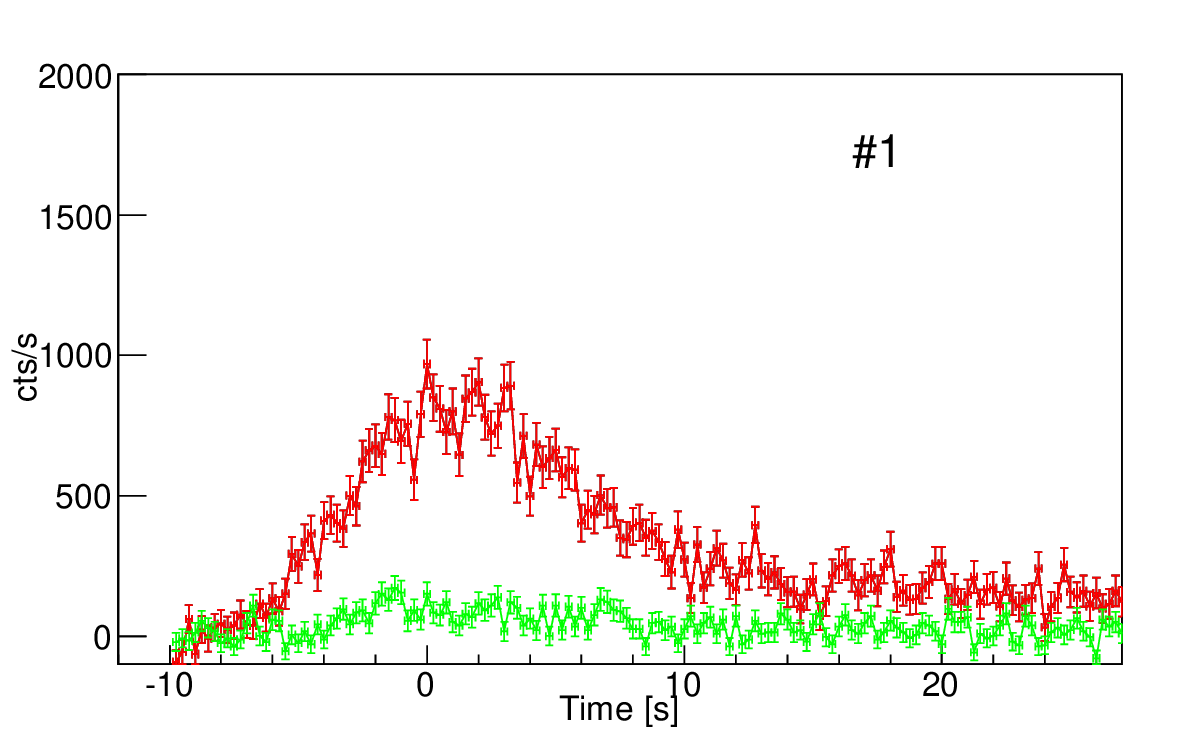}
\includegraphics[angle=0, scale=0.2]{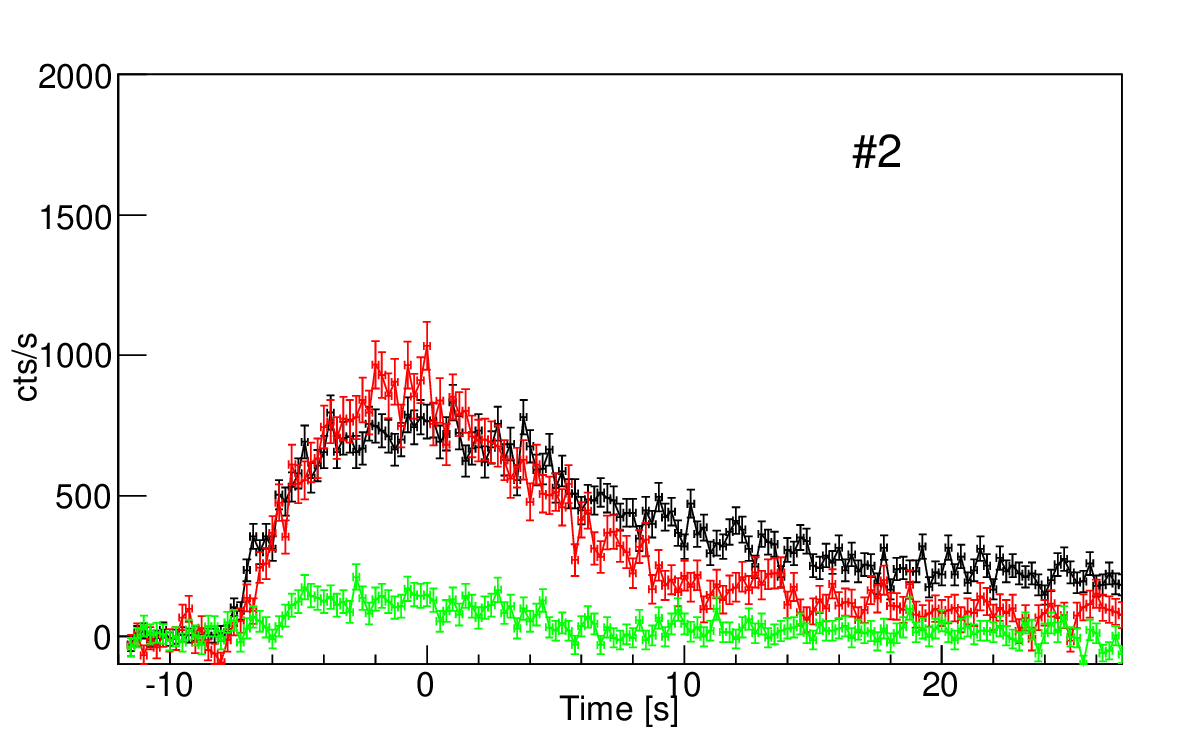}
\includegraphics[angle=0, scale=0.2]{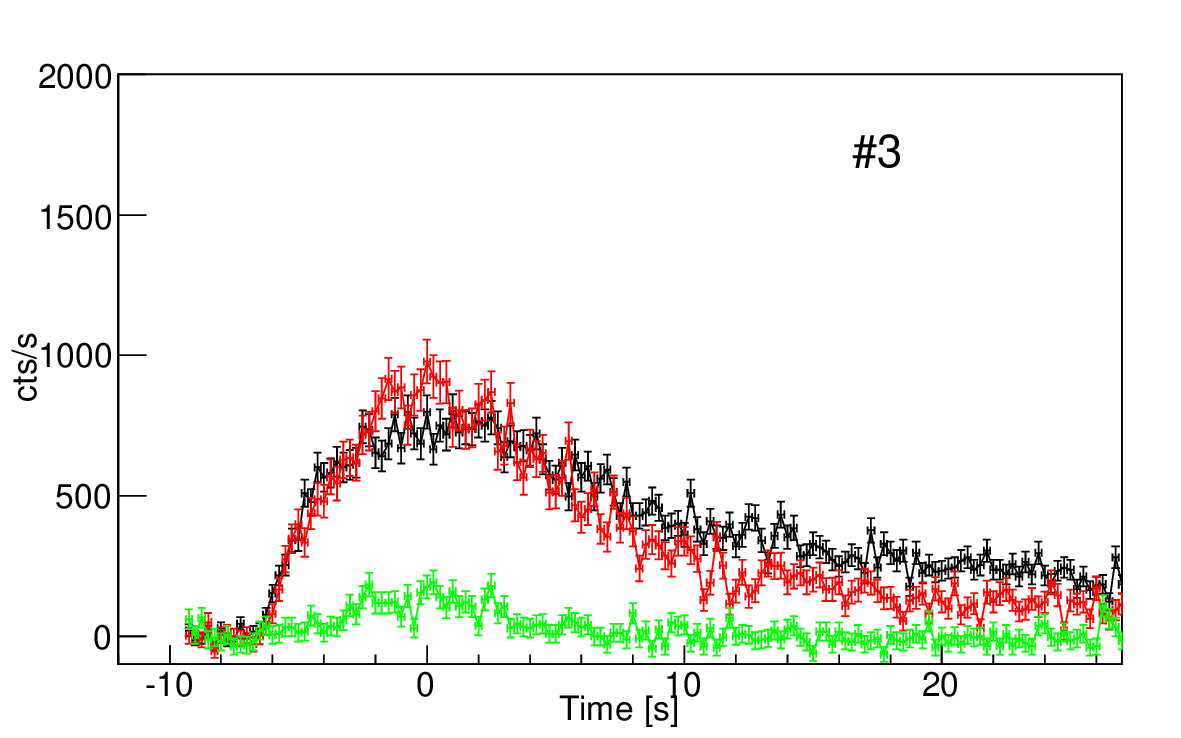}
\includegraphics[angle=0, scale=0.2]{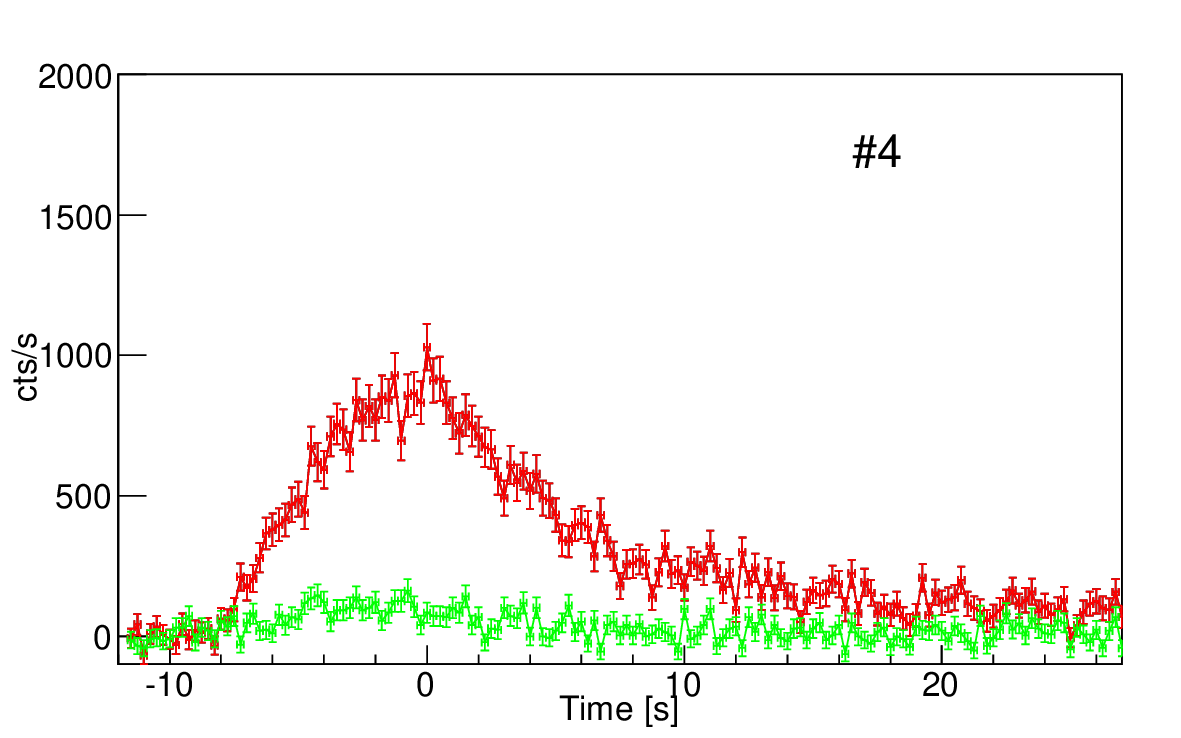}
\includegraphics[angle=0, scale=0.2]{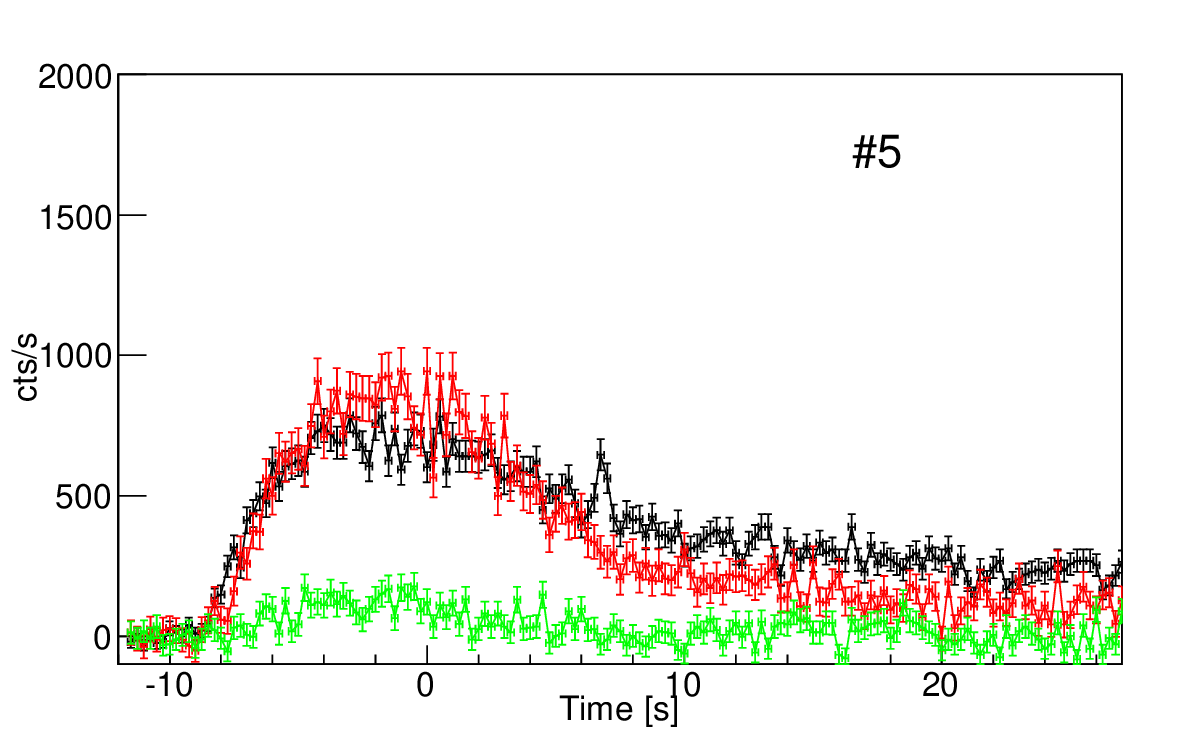}
\includegraphics[angle=0, scale=0.2]{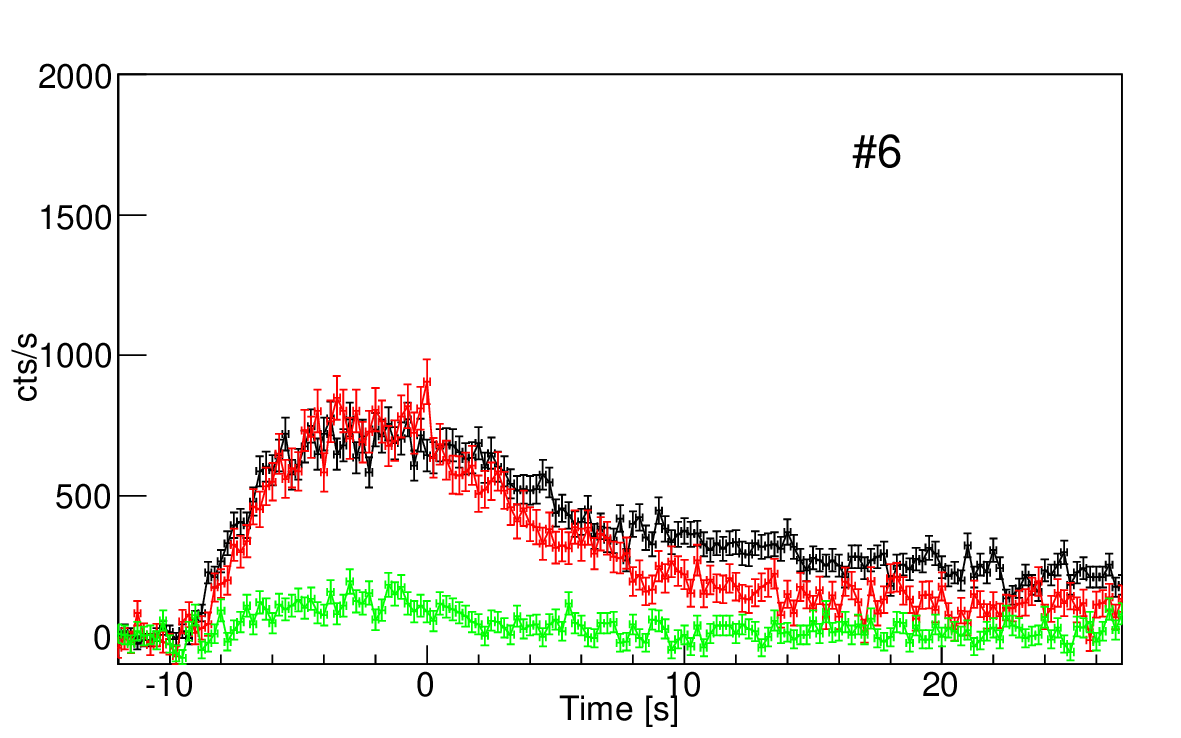}
\includegraphics[angle=0, scale=0.2]{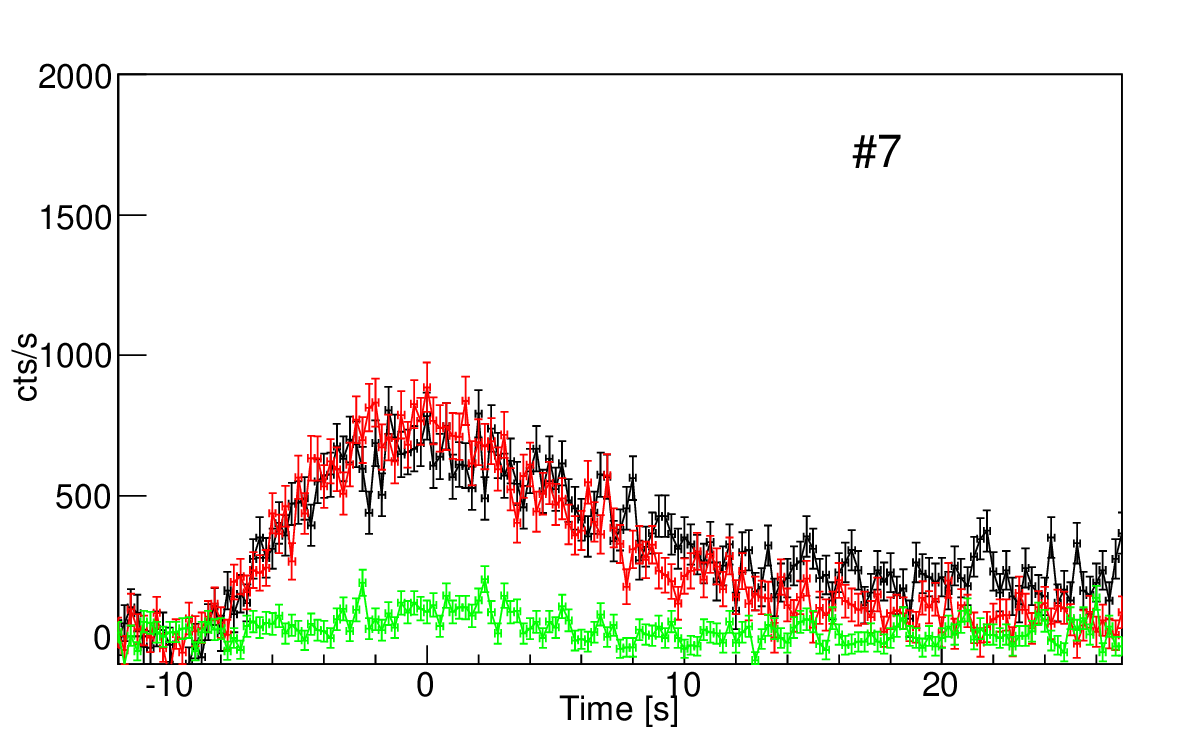}
\includegraphics[angle=0, scale=0.2]{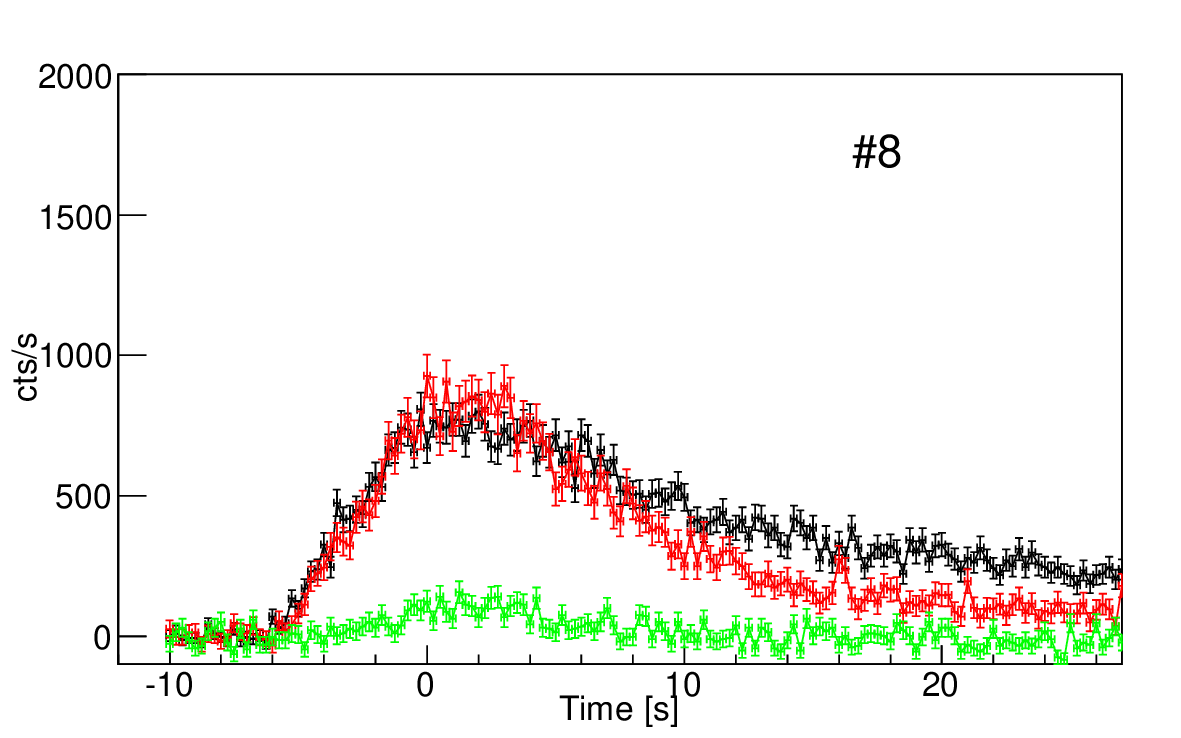}
\includegraphics[angle=0, scale=0.2]{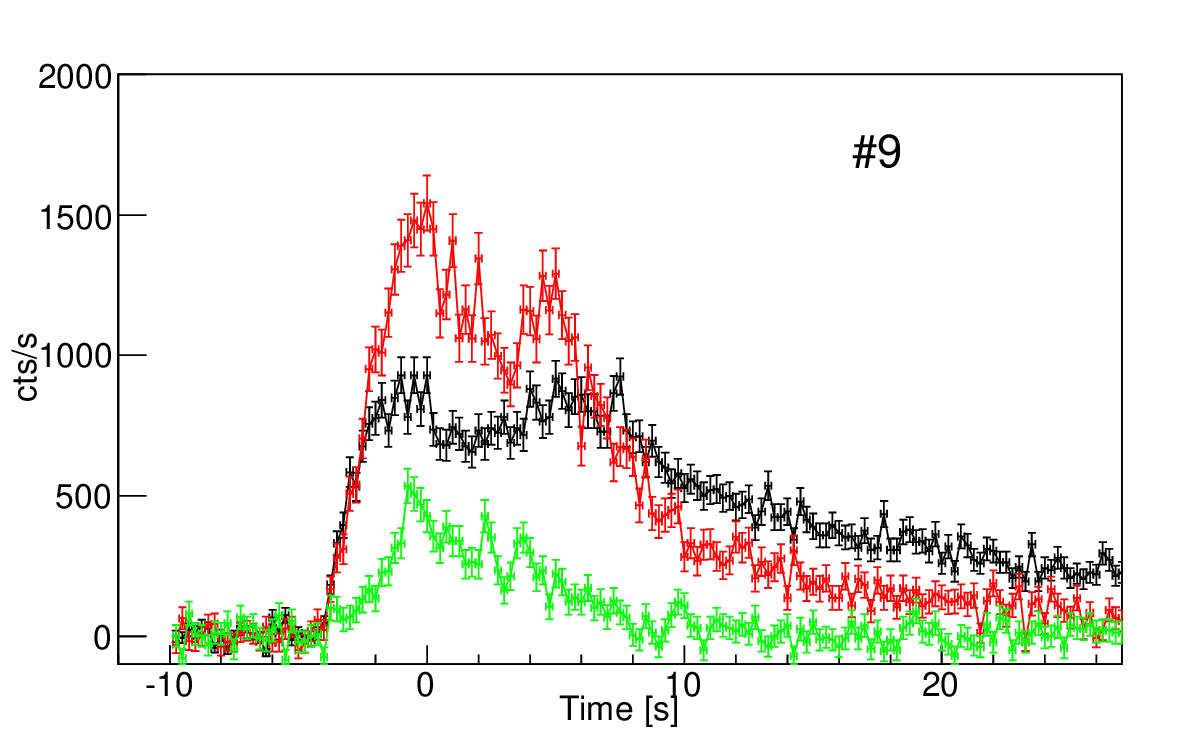}
\includegraphics[angle=0, scale=0.2]{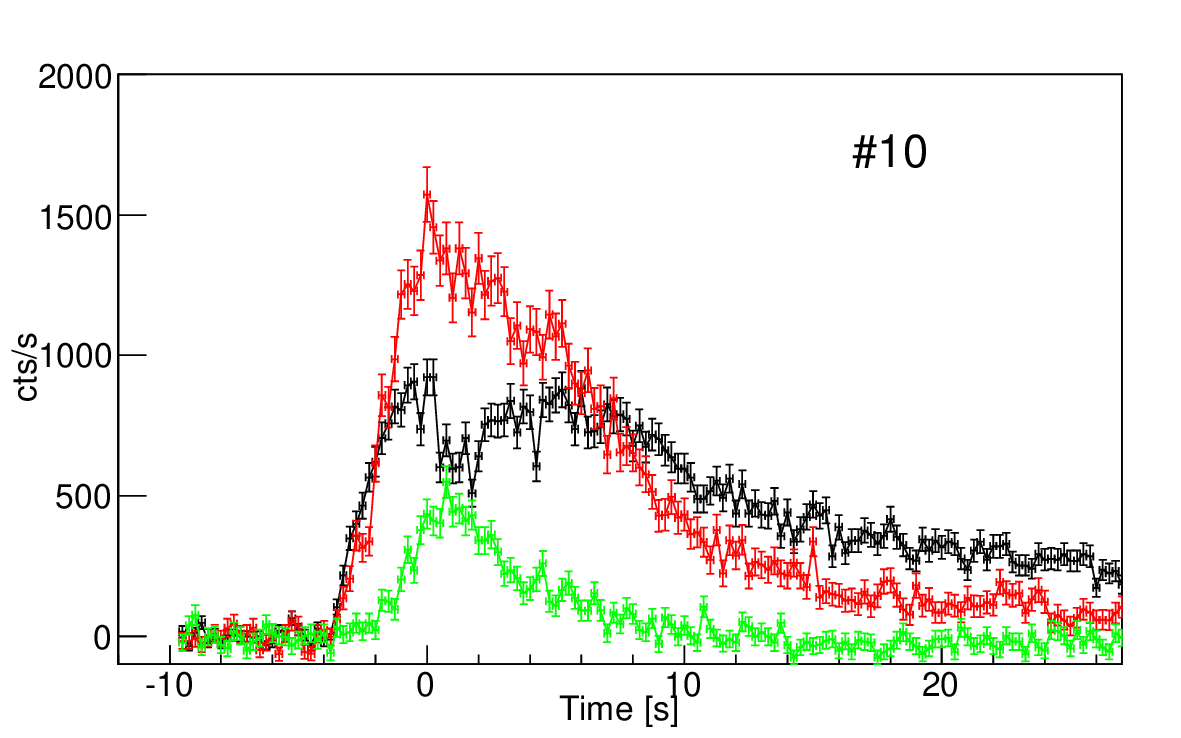}
 \caption{
 Lightcurves with pre-burst emission subtracted of the 10 type-I X-ray bursts detected in the Insight-HXMT observation of 4U~1608--52 with time bin 0.25 s by LE (black), ME (red) and HE (green). The lightcurves of LE and ME are in  their full energy bands; the energy band of HE lightcurves is  20--50 keV.
  % Background is subtracted.
  }
\label{fig_burst_lc}
\end{figure}

To avoid missing any burst from these observations, the data {\bf products} were extracted from the unfiltered data.
In practice, the good-time interval (GTI) files were extracted without any filtering, which were used for extracting screened events, lightcures and spectra. %as we did in the work on MAXI~J1816--195.
As shown in Table \ref{tb_burst} and Figure \ref{fig_burst_lc}, 10 bursts were found in the ME and HE data, but 2 of them lacked the LE data.
To catch the burst spectral evolution and achieve a balance between as small as possible of the exposure time and enough photons of each slice of the burst,
we used 0.5 s as a fixed exposure time to dissect the burst from the onset of each burst  (defined as the time 10 s before the burst peak) and took the burst peak time of the ME lightcurve as the time 0 s.
As a conventional procedure, the spectrum of the pre-burst (including the  persistent emission and the instrument background) with the time window between 70 s and 20 s before the burst peak was extracted as the background when fitting the burst spectrum.

For the spectral fitting of the persistent emission, the energy bands for LE, ME, and HE are 1.5--7 keV, 8--30 keV, and 28--100 keV, respectively. For the  spectral fitting {\bf of the burst emission}, the energy bands  for LE, ME, and HE  are 1--10 keV, 8--30 keV and 28--50 keV, respectively. The reduced energy bands of LE and HE in the persistent spectral fitting are due to the constraint of the background model.
%The  shrunken energy band of HE in the burst spectral fitting is due to the low flux level (even negative flux level) of the burst emission above 50 keV.

In the spectral fitting of both the persistent and burst emissions, the {\sc tbabs} model was adopted to account for the photoionization absorption by the interstellar medium,
%the hydrogen column (tbabs in XSPEC) accounts for both the line-of-sight column density,
%as well as any intrinsic absorption near the source, and it
where the equivalent hydrogen column was fixed at 1.33$\times 10^{22}~{\rm cm}^{-2}$ \citep{Chen16081}.
%both for the persistent and bursts spectral fitting.
Normalization constants were included during fittings to address inter-calibration issues between instruments. {\bf
%We keep the normalization factor of the LE data with respect to the ME \& ME data to unity.
%During the spectral fitting of the persistent and burst emission, we set the normalization factor of the LE data to unity. For the persistent emission, the constants of the ME and HE data are variable during spectral fitting, while for the burst emission, the constants of the ME and HE data are fixed at unity.
When fitting the persistent emission, we set the constant of LE to unity and those of ME and HE were variable.
On the other hand, when fitting the burst emission, the constants of ME and HE data were also fixed at unity.
%We fixed the normalization factor of the LE data at  unity during spectral fitting of the persistent and burst emission. The constants of the ME and HE data are variable during spectral fitting of the persistent emission but fixed at unity during spectral fitting of the burst emission.
%For the persistent emission, the constants of the ME and HE are variable during spectral fitting;  for the burst emission, the constants of the ME and HE are fixed at unity during spectral fitting.
}
%but fixed at unity for the burst spectral fitting.
The systematic error considered was 1 percent. Uncertainties were reported at the 68\% confidence level unless noted otherwise.

\section{Results}
\subsection{Outburst/persistent emission}
The flux level in the Swift/BAT lightcurve during Insight-HXMT observations in this work is 100--50 mCrab, high enough to study the hard X-ray emission during the bursts. Moreover, the CCD depicts the outburst evolution and
{\bf indicates that bursts considered in this paper were located in the low/hard (island) state.
}
%indicates that  these observations of this work located at the
We also notice that before the high/soft state, there is a preceding low/hard state around MJD 59770  with a higher flux level in the hard X-ray band, $\sim$200 mCrab in the Swift/BAT lightcurve.
This reminds us of the two low/hard states in another burster--IGR~J17473-2721 \citep{Chen2012}, e.g., the brighter preceding low/hard state and the fainter lagging low/hard state.
For observations of this work, the outburst stays in the  fainter lagging low/hard state.

The spectra of the obsid where the burst is located, or the nearby obsid within half a day if the former obsid has a short GTI, are considered as the persistent emission since the outburst evolved with a timescale of several days, which are used to probe the accretion circumstance. At first, these spectra were fitted by an absorbed convolution thermal Comptonization model (with an input seed photon spectra blackbody), available as thcomp (a more accurate version of nthcomp; Zdziarski et al. 2020) in XSPEC, i.e., the model is tbabs*thcomp*bb, which is described by the optical depth $\tau$, electron temperature $kT_{\rm e}$, and scattered/covering fraction $f_{\rm sc}$.
{\bf Considering the source was in a low/hard state during observations, we first attempted the scenario of a spherical corona}.
However, the derived blackbody temperatures and radii $T_{\rm bb}$ were found to be 0.30$\pm$0.03--0.45 $\pm$0.03 keV and 36$\pm$7km--20$\pm$2km, which is unlikely.
Then the scenario of a disk corona was employed, i.e., the model is revised to tbabs*thcomp*diskbb\footnote{{\bf It is possible that the pre-burst/persistent spectra are well described by the disk-corona scenario and lack the quasi-spherical corona, which differs from the corona geometry derived from the burst hard X-ray deficit. This discrepancy could be due to the data quality issues, e.g., lack of  photons below 1.5 keV and low count rate due to the small effective area of LE.}}.
In this case, the derived parameters are shown in Table \ref{tb_fit_thcomp_diskbb}.
The inferred bolometric fluxes are corresponding 5.9\%--3.8\% $L_{\rm Edd}$ at distance of 4 kpc where $L_{\rm Edd}=1.8\times10^{38}$ erg/s.

\subsection{Burst lightcurves}
Based on the LE/ME/HE lightcurves (Figure \ref{fig_burst_lc}) with a time resolution of 0.25 s and the bursts' parameters (Table \ref{tb_burst}), it is evident that the first 8 bursts share similar characteristics in terms of burst profile, fluence, peak flux, and duration. In contrast, the last 2 bursts exhibit distinct features, being brighter and shorter than the preceding 8 bursts. Additionally, the last 2 bursts display double-peaked structures in the LE or ME lightcurves, setting them apart from the earlier bursts. Consequently, these bursts are being investigated as two separate groups.

\begin{figure}
\centering
\includegraphics[angle=0, scale=0.5]{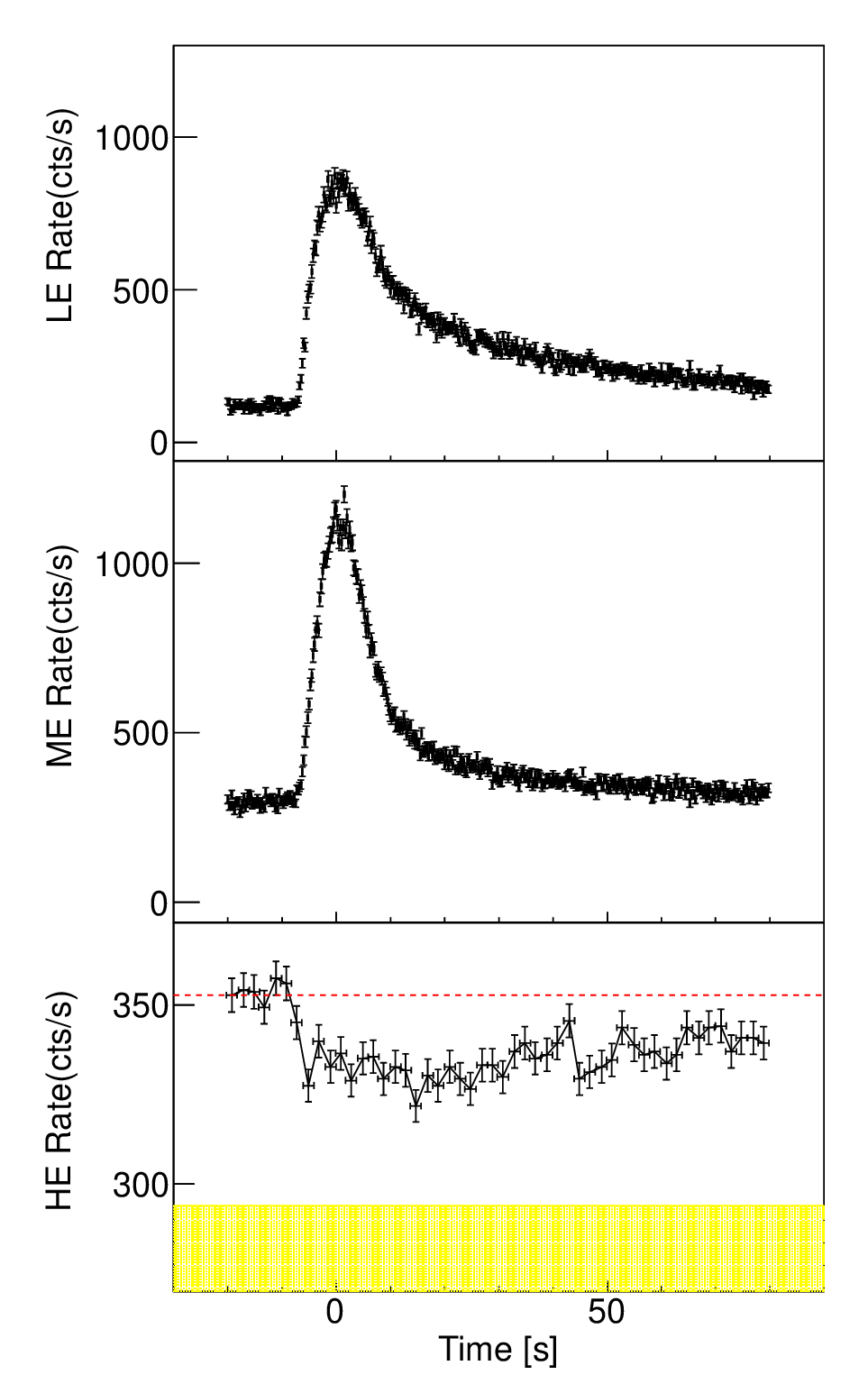}
\includegraphics[angle=0, scale=0.5]{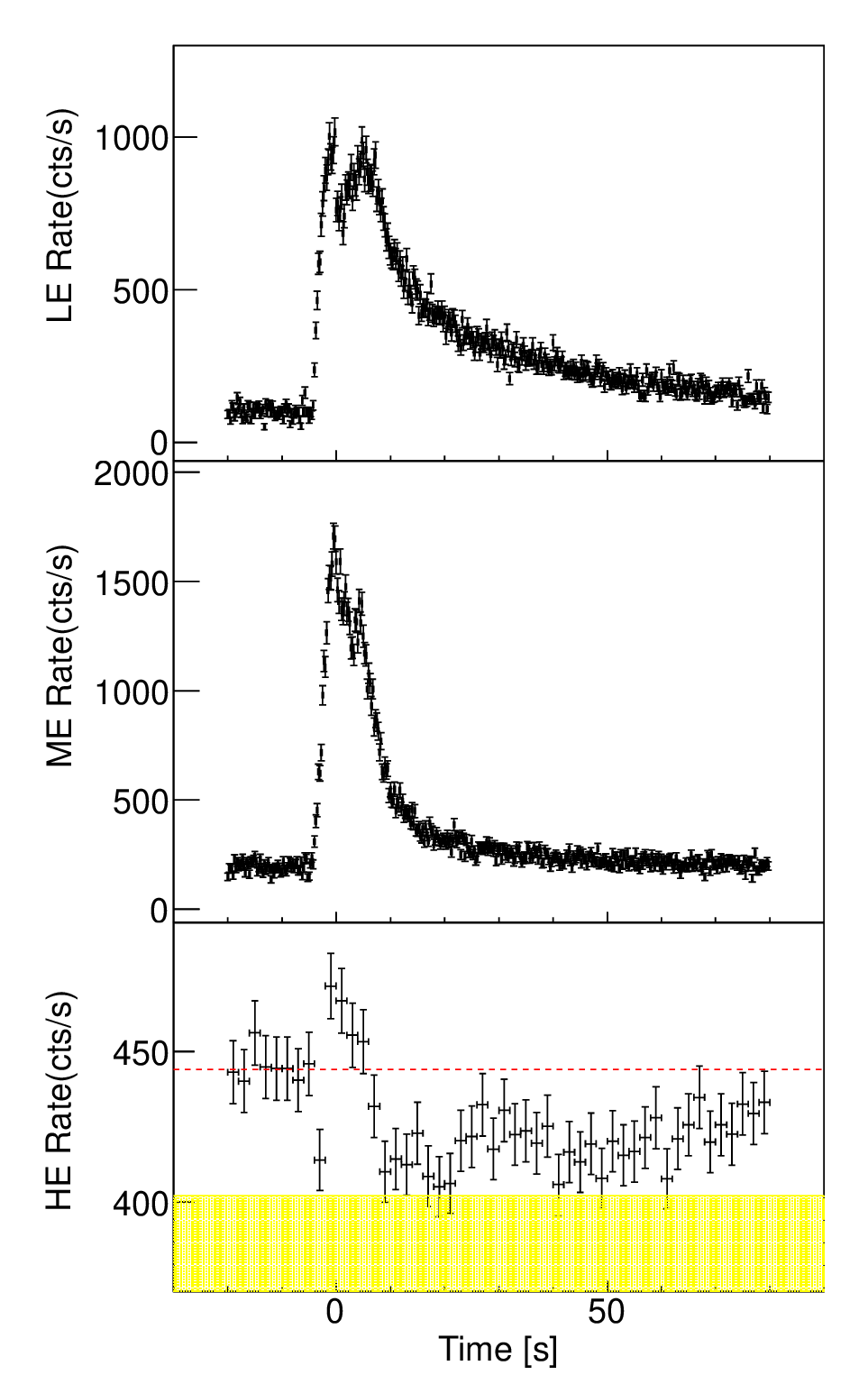}
  \caption{For the first 8 bursts (left panels) and last 2 bursts (right panels), the stacked lightcurves of LE (top), ME (middle), and HE (bottom) of the burst in 1--10 keV, 8--30 keV and 30--100 keV, respectively. The time bin for LE and ME is 0.25 s, for the HE is 2 s. The red lines and the yellow zones in the bottom panels indicate the pre-burst emission (persistent emission and background) and  the background level for the HE detectors, respectively.   }
  \label{fig_burst_lc_stacked}
\end{figure}

In order to study the hard X-ray lightcurves during the bursts, burst profiles in the 30--100 keV bands were generated by HE, revealing a hard X-ray deficit. To enhance statistical significance, the lightcurves of the first 8 bursts and the last 2 bursts were stacked separately. For comparison, the LE and ME lightcurves were also stacked. The averaged lightcurves of these two groups of bursts are presented in  Figure \ref{fig_burst_lc_stacked}.
In the bottom panel of the left panels in Figure \ref{fig_burst_lc_stacked}, a decrease in the hard X-ray band is observed, accompanying the burst rise detected by LE and ME.
The hard X-ray {\bf deficit}  weakens following the burst decay.
Considering the pre-burst emission of 353 cts/s (including $\sim$59 cts/s persistent emission and 294 cts/s background emission), the HE decrement reaches a maximum of approximately 30 cts/s at the burst peak,  amounting to $\sim$ 50\% of the whole persistent flux in 30--100 keV.
The average deficit from $-$10 s to 40 s (burst peak time as 0 s) is $-19.9\pm0.9$ cts/s.

A cross-correlation analysis is performed between LE and HE lightcurves with a bin size of 0.25\,s, following the method outlined by \citet{Chen2012}. The minimum of the cross-correlation function appears at $\sim$ 1\,s, which indicates that the hard X-ray deficit lags the burst emission by $\sim$ 1\,s.

For the last 2 bursts, as shown in the right panels of Figure \ref{fig_burst_lc_stacked},
a flux-rising is observed accompanying the burst rise detected by LE and ME, followed by a flux-dropping after the burst peak in the HE lightcurves.
Following the decay of the bursts, the HE flux returns to the pre-burst level.
Considering the pre-burst emission of 440 cts/s (including $\sim$38 cts/s persistent emission and 402 cts/s background emission), the HE decrement reaches a maximum of 40 cts/s at the burst peak, which amounts to $\sim$ 100\% of the whole persistent flux in 30--100 keV. However, the larger error bars of the deficit prevent a conclusive determination of the maximum deficit degree.
The average deficit from 10 s to 60 s (burst peak time as 0 s) is $-22.2\pm2.0$ cts/s.

\subsection{Time-Resolved Spectroscopy of burst emission}

\begin{figure}
\centering
\includegraphics[angle=0, scale=0.2]{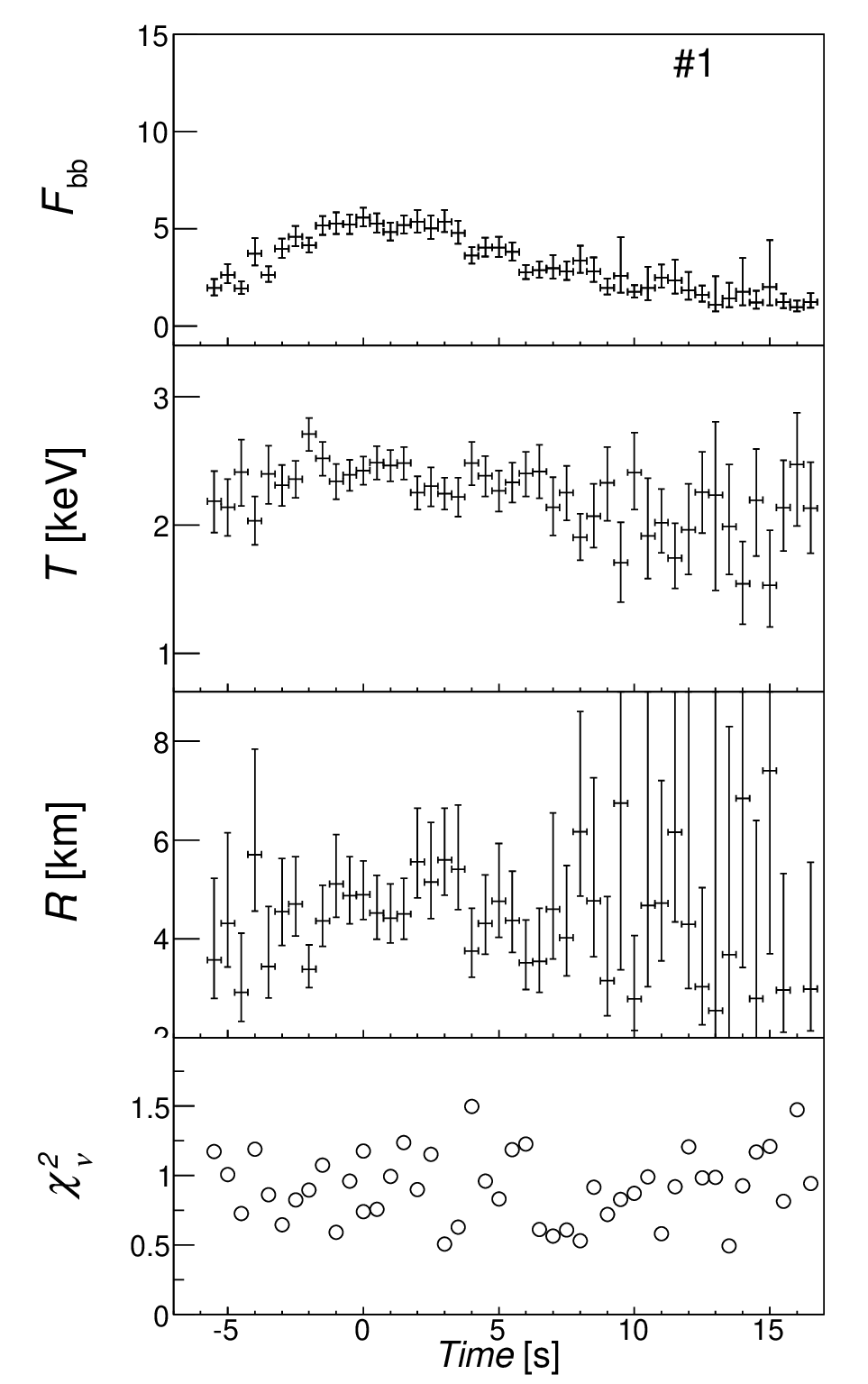}
\includegraphics[angle=0, scale=0.2]{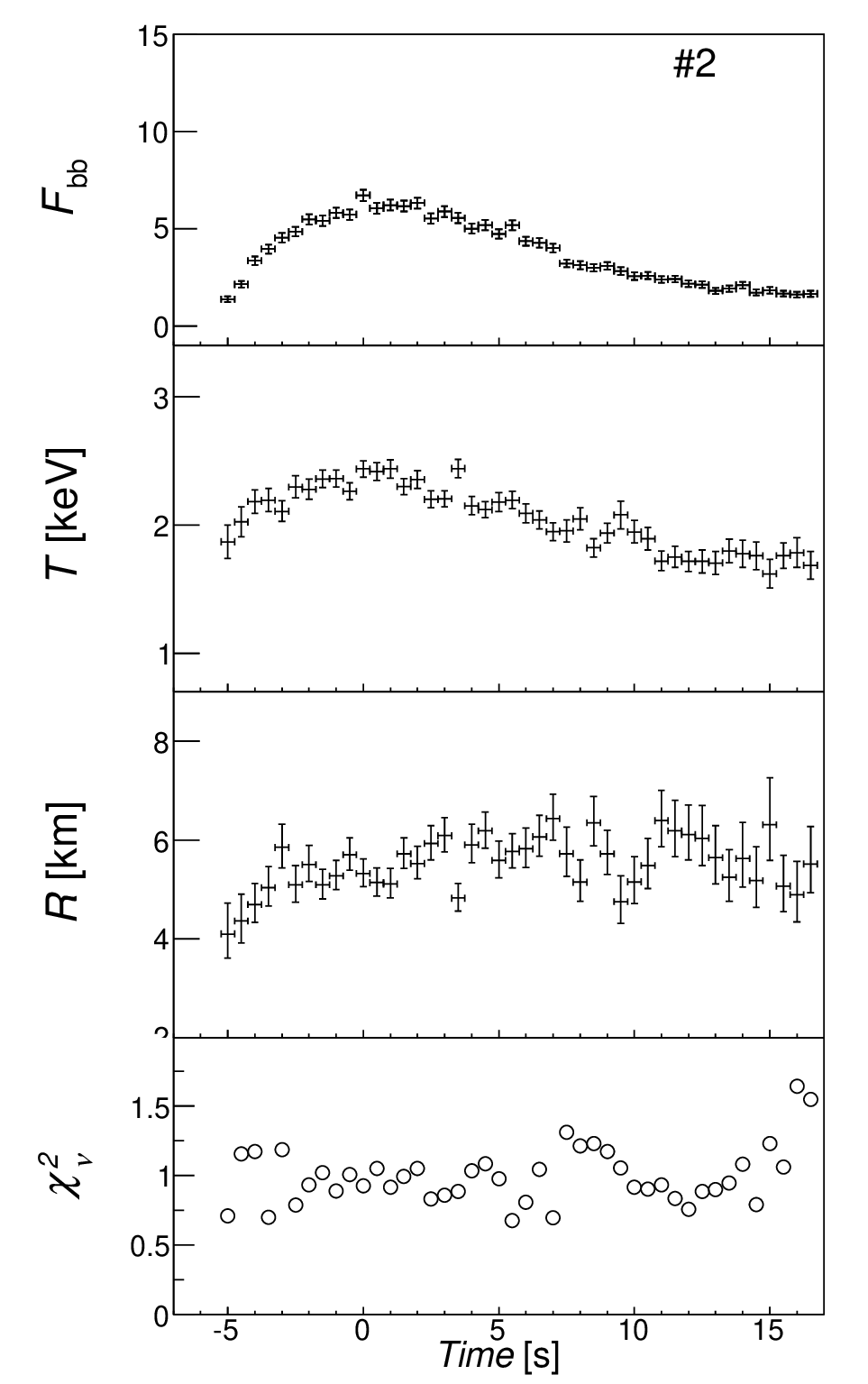}
\includegraphics[angle=0, scale=0.2]{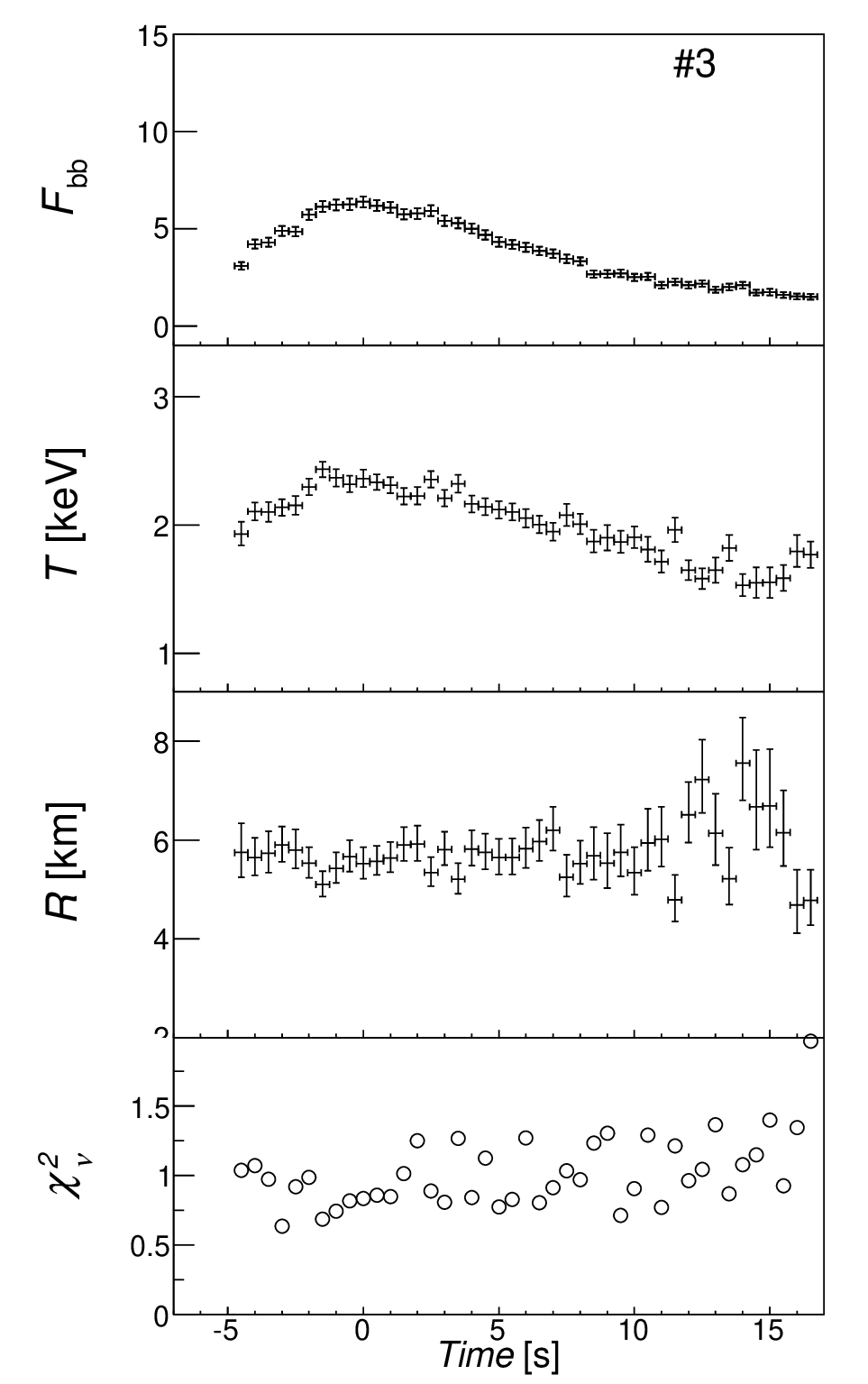}
\includegraphics[angle=0, scale=0.2]{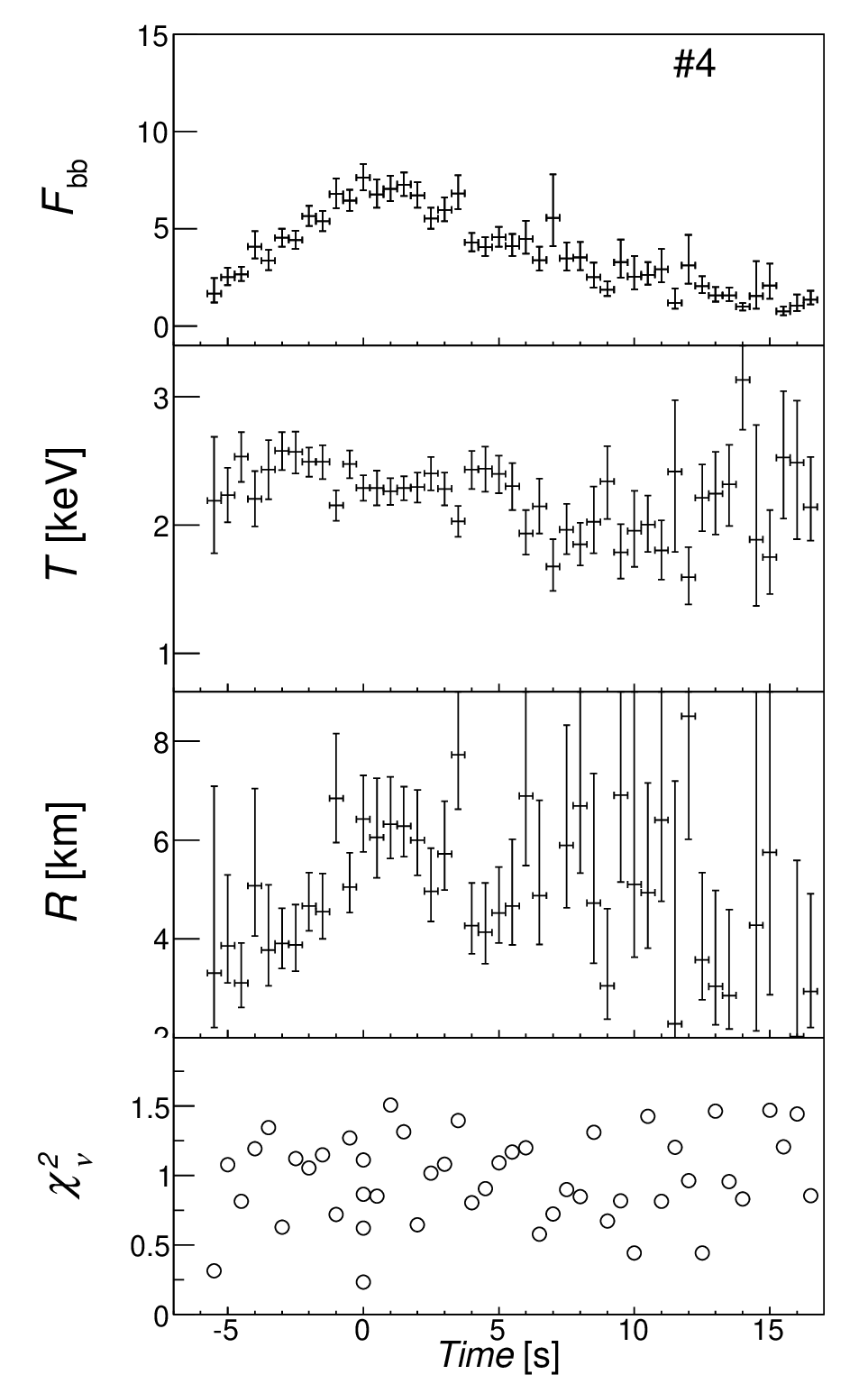}
\includegraphics[angle=0, scale=0.2]{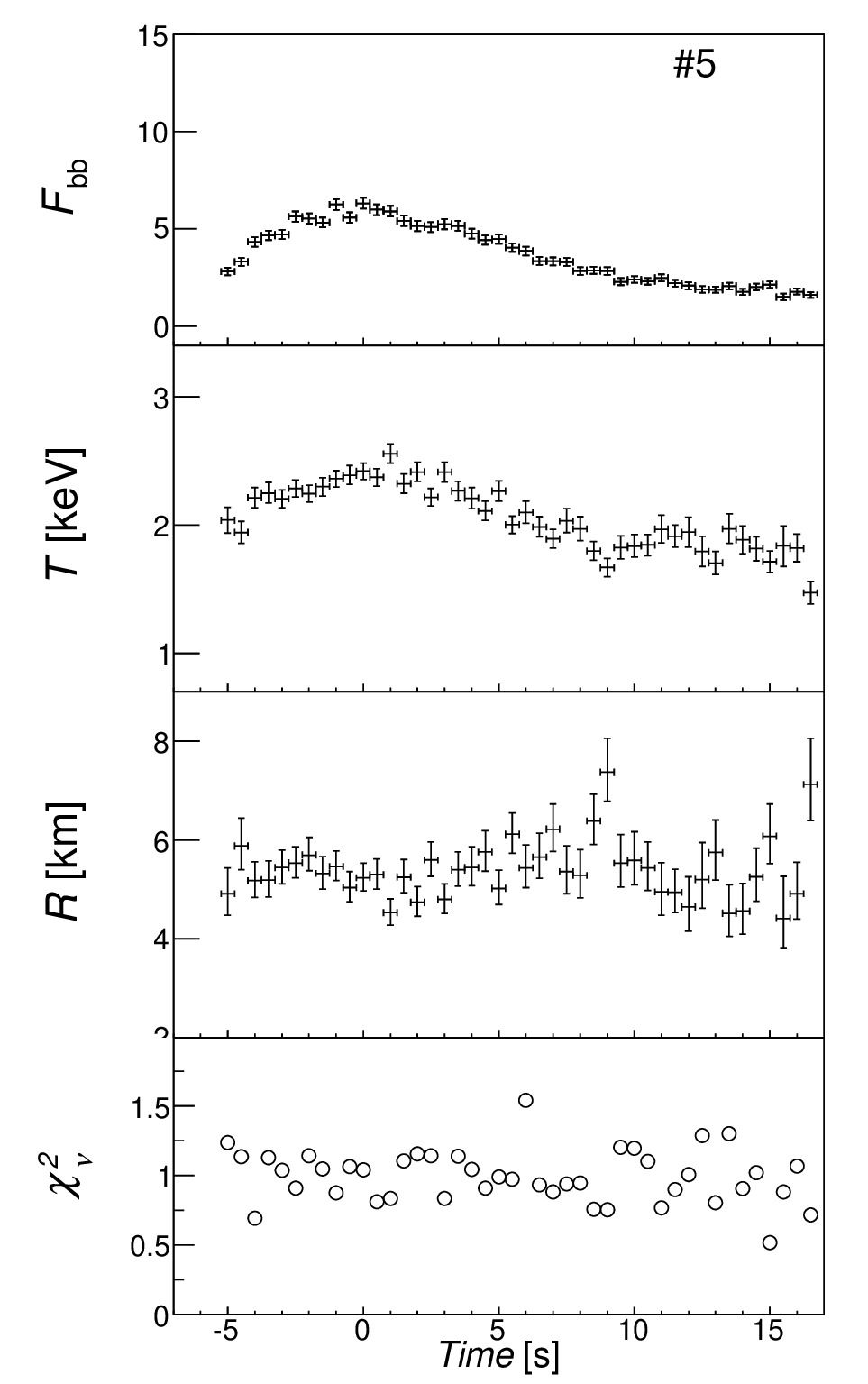}
\includegraphics[angle=0, scale=0.2]{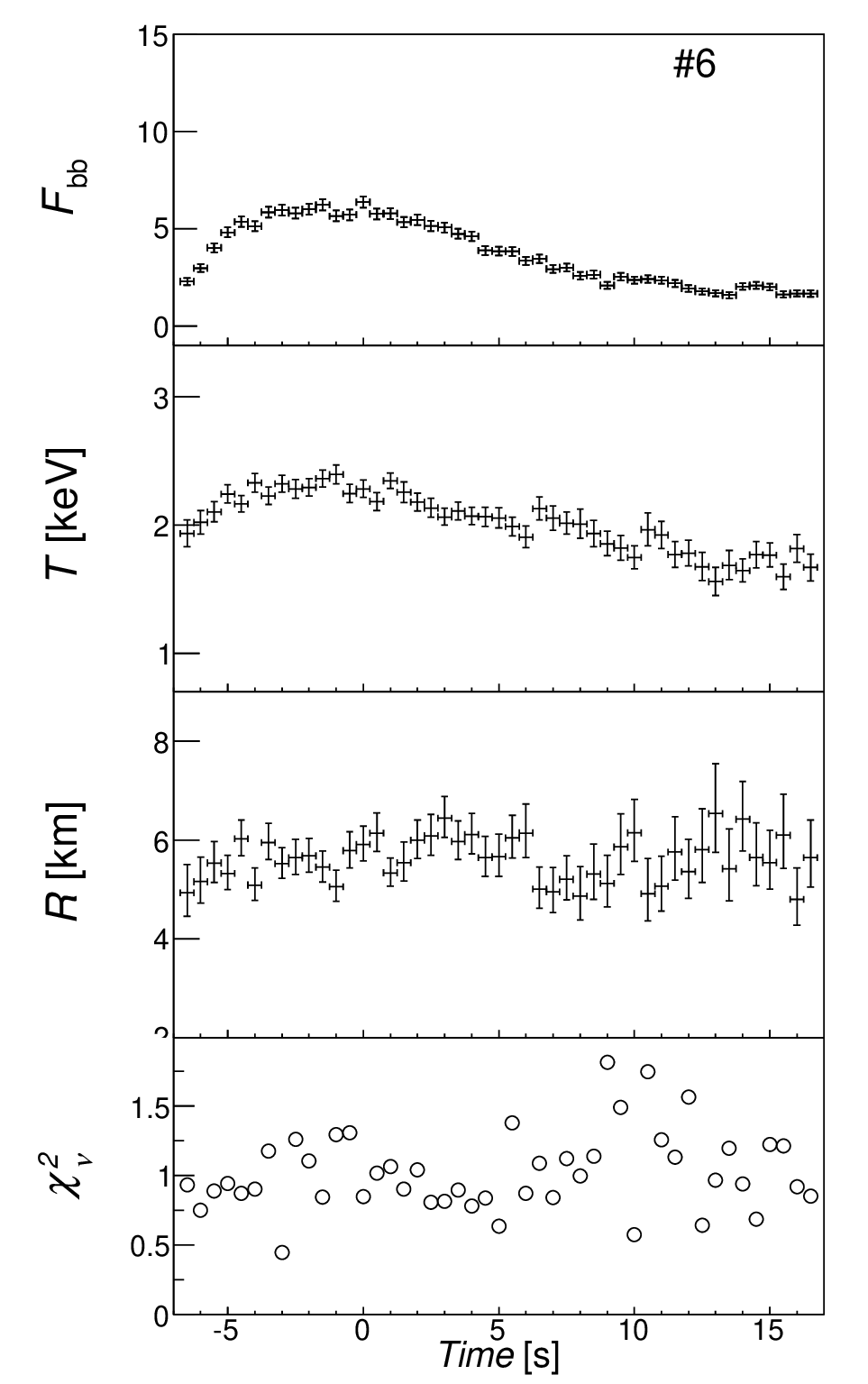}
\includegraphics[angle=0, scale=0.2]{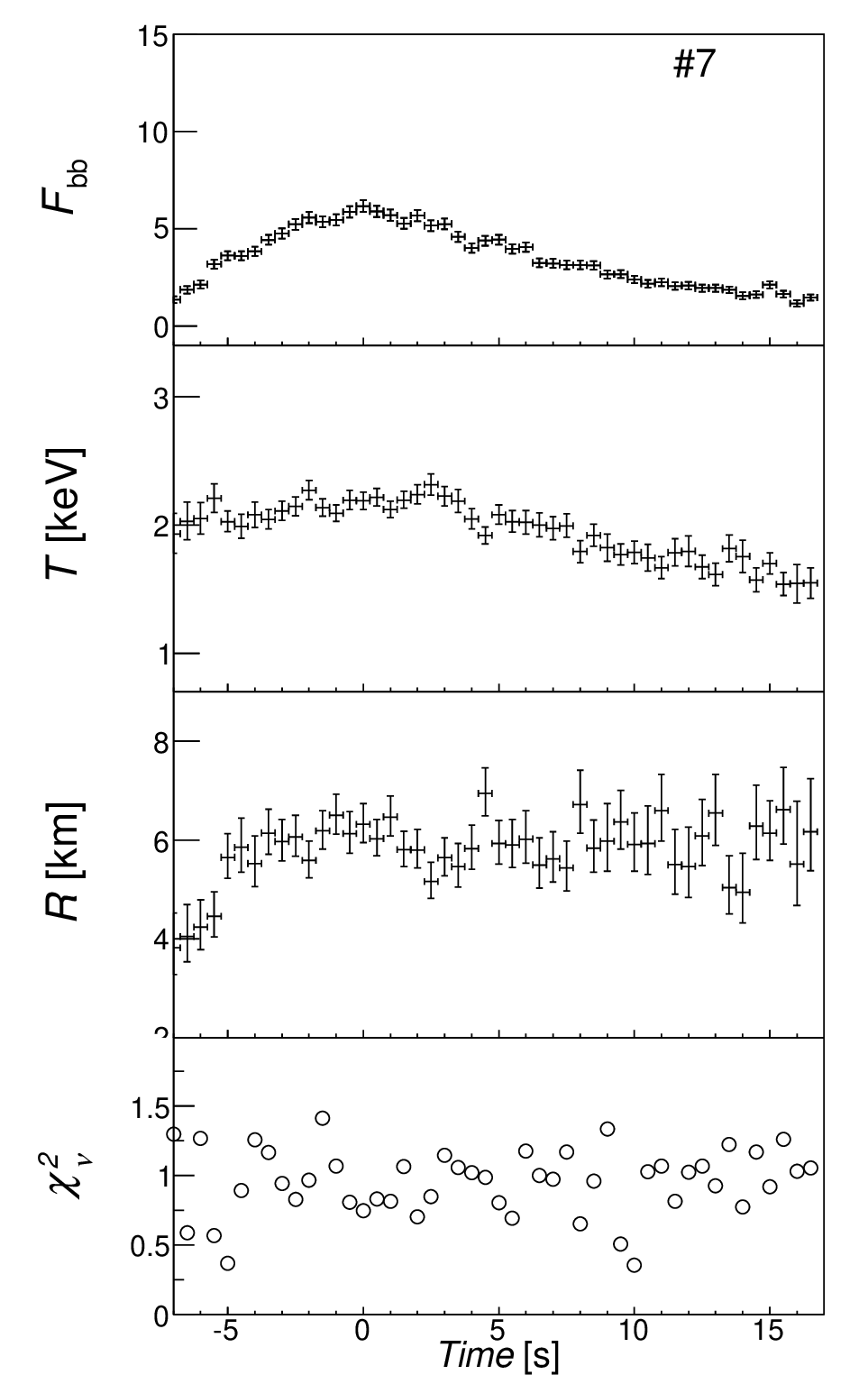}
\includegraphics[angle=0, scale=0.2]{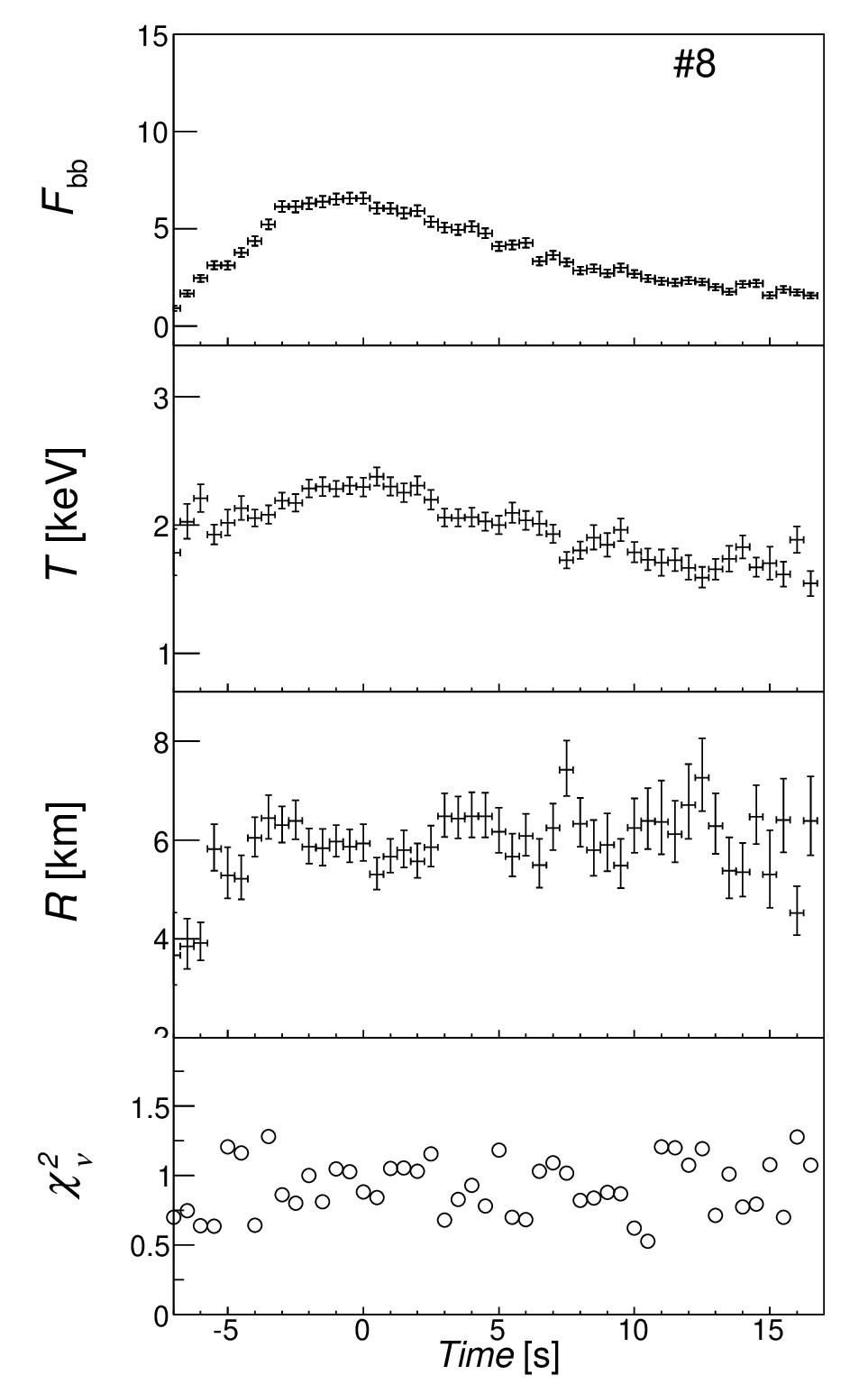}
\includegraphics[angle=0, scale=0.2]{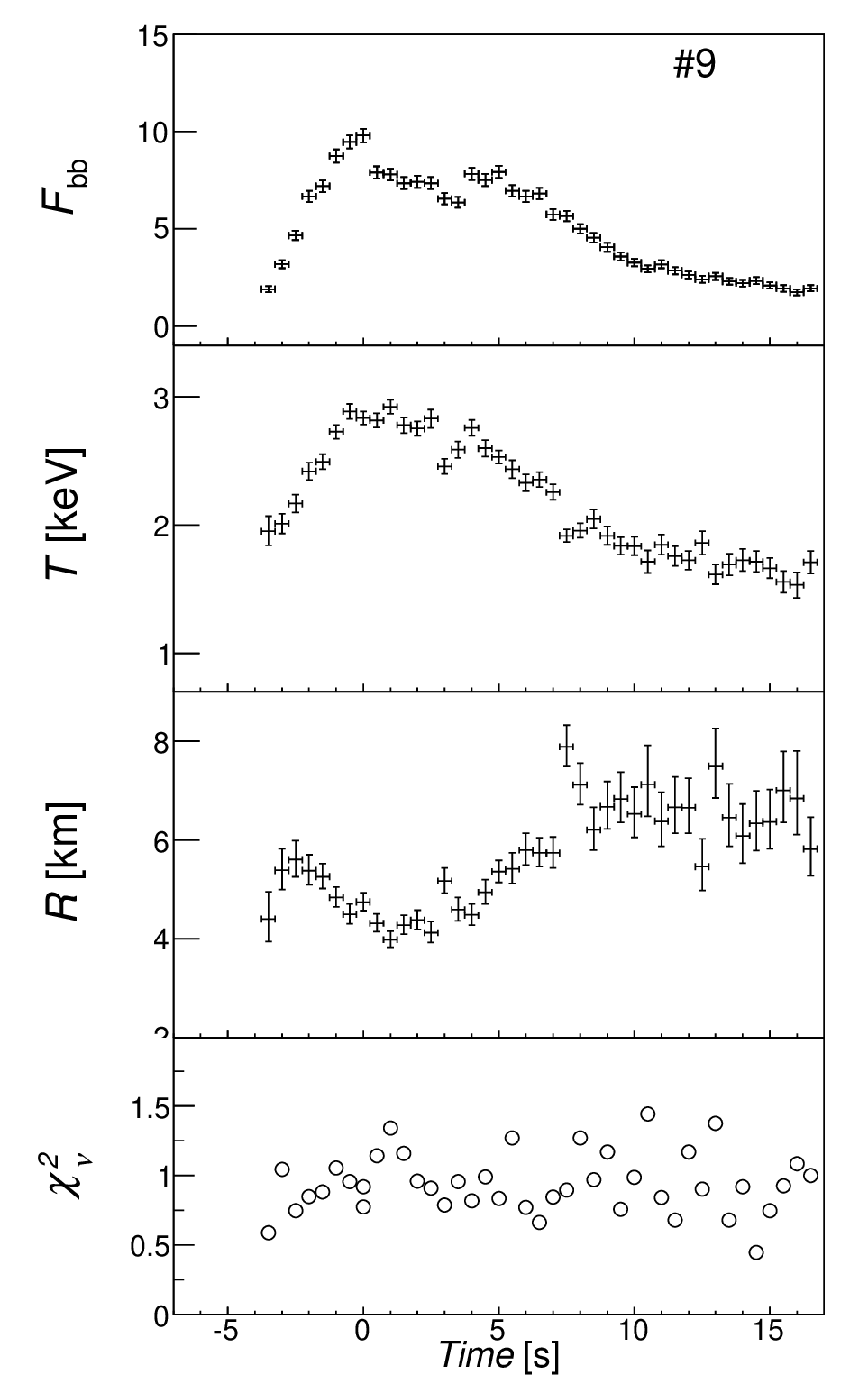}
\includegraphics[angle=0, scale=0.2]{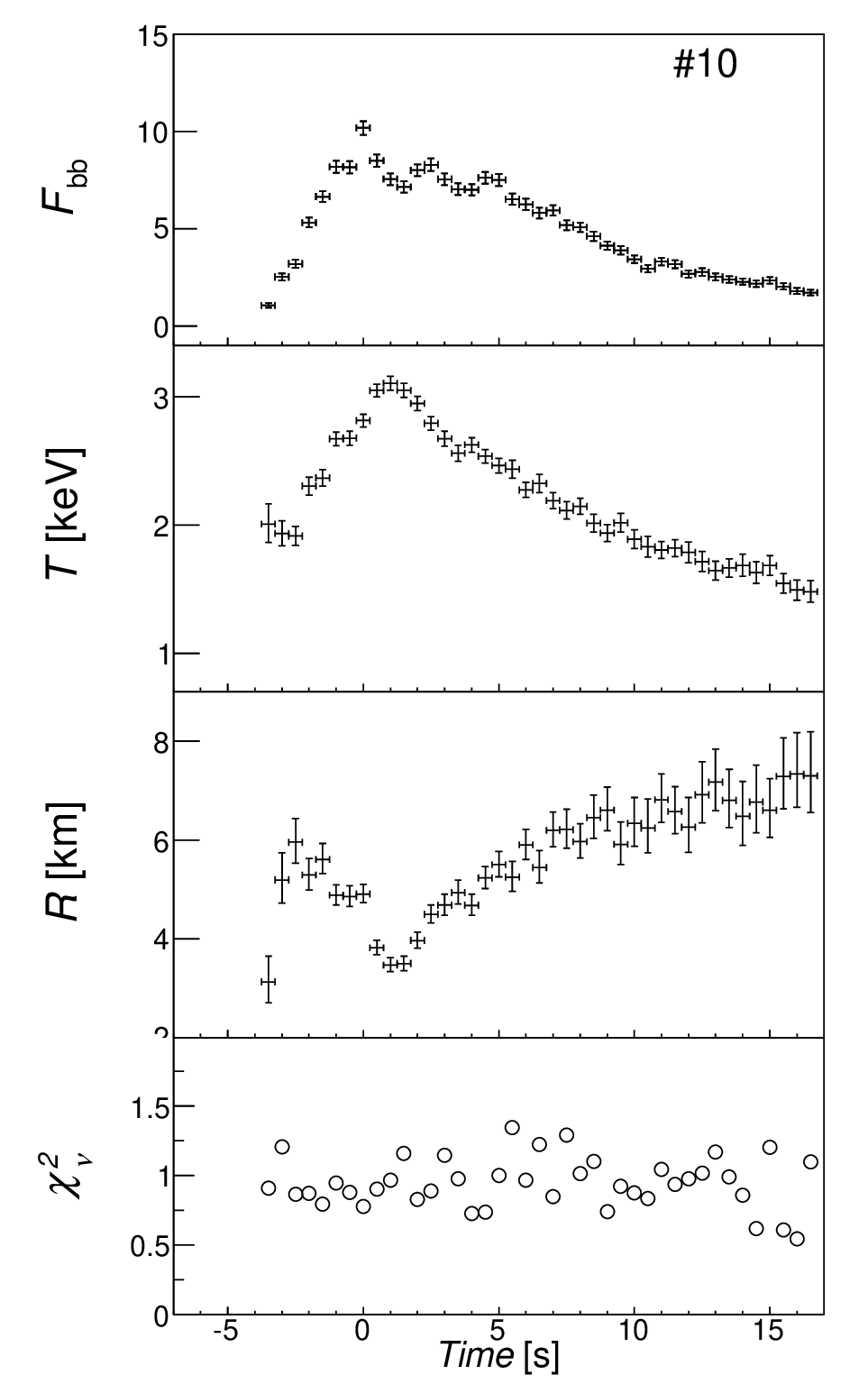}
\caption{
 Results of the spectral fits of time-resolved spectra of the 10 bursts detected from 4U~1608--52 during its  2022 outburst, including the
  blackbody bolometric flux $F_{\rm bb}$ in units of $10^{-8}~{\rm erg/cm}^{2}/{\rm s}$ , the temperature $kT_{\rm bb}$, the observed radius $R$ of the NS photosphere at 4 kpc, and the goodness of fit $\chi_{v}^{2}$.
 %The red lines indicate the fitting results of the temporal evolution of the bolometric flux   with an exponential decay function.
 %All of the bursts except the  second-to-last burst show photospheric radius expansion.
% We  measured additional decay constant(s) $\tau_{1}$ and $\tau_{2}$  for the single (bursts \#1--\#8) or broken double (bursts \#9)  exponential fits to the bolometric flux evolution.
 %; the  second-to-last burst  shows a double peak profile in flux
  % Background is subtracted.
  }
\label{fig_burst_fit_bb}
\end{figure}

We modeled the burst spectra with the conventional model--an absorbed blackbody, and the derived model parameters are shown in Figure \ref{fig_burst_fit_bb}.
The distinction between the two groups of bursts, as indicated by the lightcurves, is also reflected in the spectral evolution.
The first 8 bursts exhibit a clear similarity in spectral evolution, with all of them having a peak temperature of $\sim$ 2.3 keV, a peak radius of $\sim$ 6 km, and a slow-changing temperature and radius during the burst decay phase.
Conversely, the last two bursts display a peculiar characteristic, with a higher temperature of up to 3.1$\pm$0.5 keV, a smaller peak radius 3.5$\pm$0.14 km and an `M' shape of the radius time evolution.
Notably, the radius reaches its minimum value of 3.5$\pm$0.14 km at the burst peak time and reaches its maximum value of 6.9$\pm$1.0 km at the burst tail.

\begin{figure}
\centering
\includegraphics[angle=0, scale=0.5]{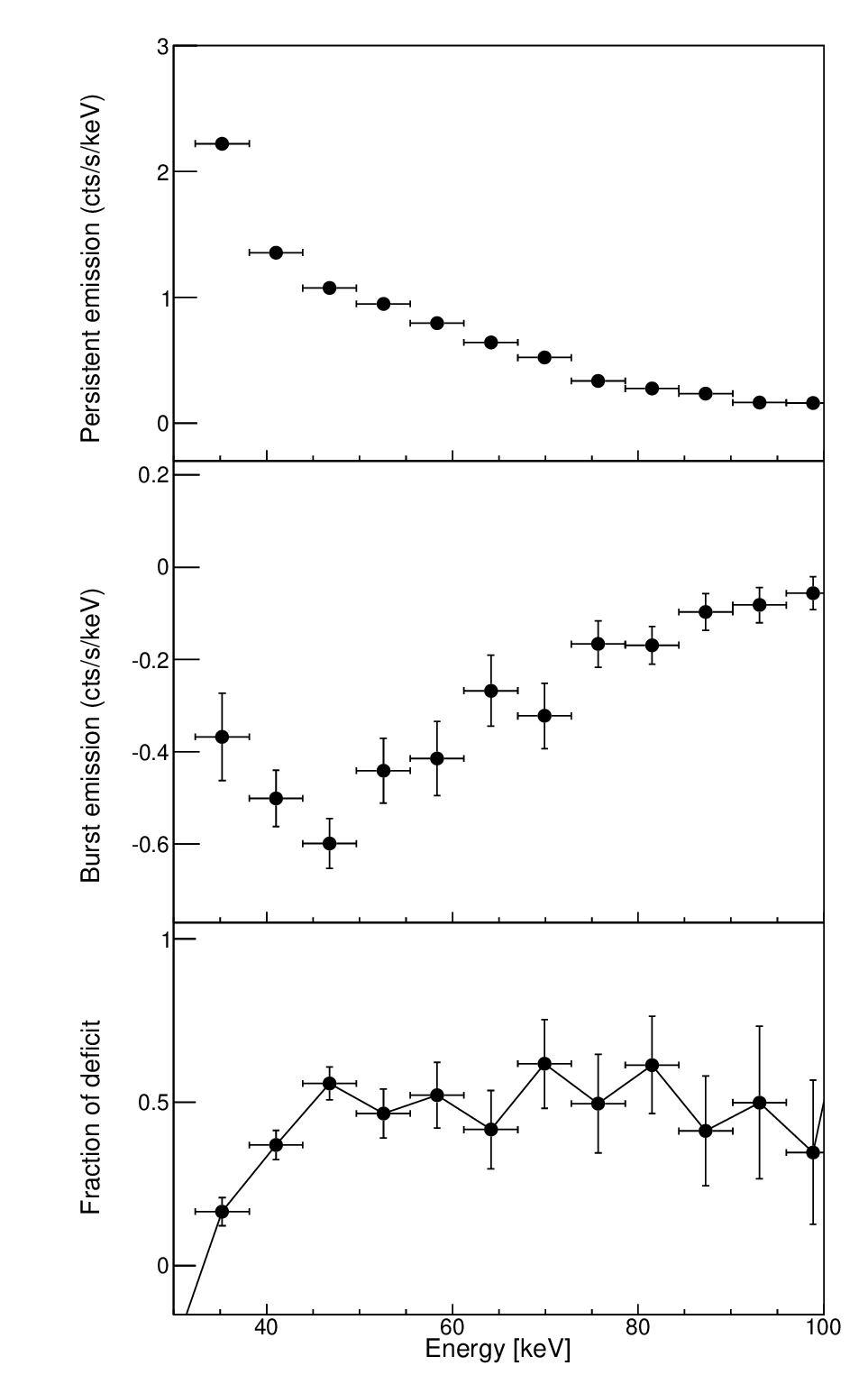}
  \caption{For the first 8 bursts  at [-10 s, 40 s], top panel: the  spectrum of the persistent emission by HE, middle panel:  the detected spectrum  of the bursts, bottom panel: deficit fraction VS energy during the bursts detected by HE.
  }
\label{fig_burst_fake}
\end{figure}

Following the methodology employed in the study of MAXI~J1816--195 \citep{Chen1816}, we investigated the variation of the hard X-ray fraction with the energy for the two groups of bursts. For each group, two spectra are used as inputs for estimation, i.e., the averaged spectrum of the persistent emission (the top panel of Figure \ref{fig_burst_fake}) and the averaged spectrum of the bursts (the middle panela of Figure \ref{fig_burst_fake}). The first spectrum was obtained by stacking the persistent spectra of the obsid during which the bursts occurred using the ftool addspec. The last spectrum was derived by stacking the spectra of the bursts emission.
For the first 8 bursts, the time interval of the burst spectra is between -5 s to 45 s, i.e., [-5 s, 45 s]. For the last 2 bursts, there is still a contribution from the bursts around its peak time in the hard X-ray band, leading to a revision of the time interval of the burst spectra to 10 s to 60 s after the burst peak time, i.e., [10 s, 60 s].
The fraction of the deficit was derived based on the two averaged spectra above,
i.e., the value of the deficit divided by the persistent emission, which is shown in Figure \ref{fig_burst_fake}.
For the first 8 bursts,  the fraction of the deficit increases with the energy in 30--50 keV and stabilizes at $\sim$50\% in 50--100 keV.
For the last 2 bursts, it is challenging to ascertain a robust evolution of the fraction of the deficit versus the energy due to the low detection significance of the hard X-ray deficit.

\section{Discussion}

{\bf Inspired} by the evolution of the hard X-ray deficit fraction with energy  detected in MAXI~J1816--195, Insight-HXMT performed cadence observations in the low/hard state of 4U~1608--52 and detected 10 thermonuclear X-ray bursts in September 2022. These burst samples are divided into two groups based on their characteristics and demonstrate a discernible trend of the hard X-ray deficit fraction with energy.

\subsection{Non-PRE double-peaked  bursts}

\begin{figure}
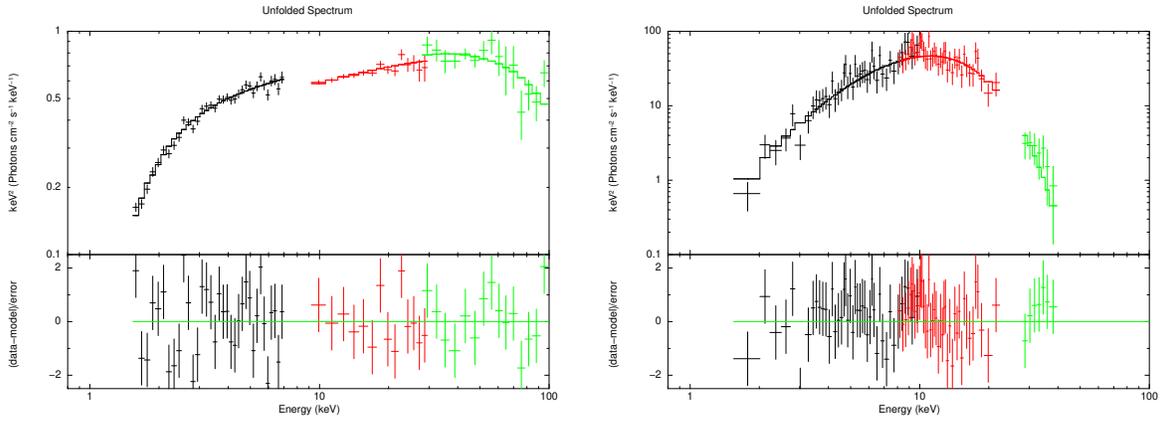

\centering
\includegraphics[angle=-90, scale=0.3]{persist_101_2.eps}
\includegraphics[angle=-90, scale=0.3]{burst_1102_10.eps}
  \caption{Left panel:The spectrum of 4U~1608--52  of Insight-HXMT/LE (black), Insight-HXMT/ME (red) and Insight-HXMT/HE (green); the best-fitting model consists of an absorbed convolution
 thermal Comptonization model (with an input seed photon spectrum diskbb).
 %Right panel: the  spectral fitting results by NICER (black), Insight-HXMT/LE (red) and Insight-HXMT/ME when the burst reached its peak flux.
 Right panel: the  spectral fitting results by Insight/LE (black), Insight-HXMT/ME (red) and Insight-HXMT/HE (green) when the burst reached its peak flux with the blackbody model.
 }
\label{fig_outburst_spec}
\end{figure}

\begin{figure}
\centering
\includegraphics[angle=0, scale=0.5]{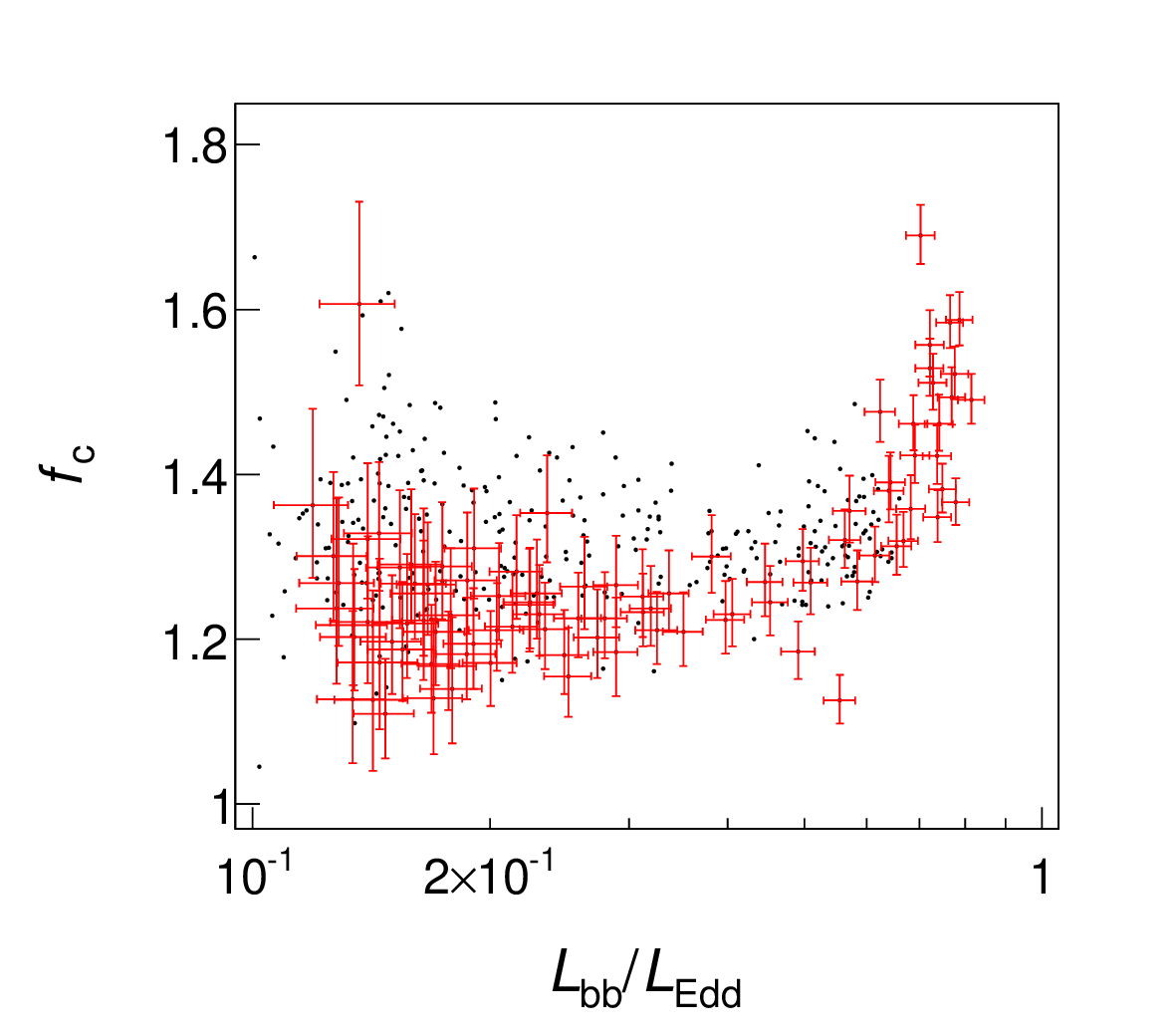}
  \caption{
  The dependence $f_{\rm c}$--$L_{\rm bb}/L_{\rm Edd}$ as observed during the cooling track of the first 8 bursts (black) and last 2 bursts (red) from 4U~1608--52 by assuming that the NS radius is 10 km and $L_{\rm Edd}$ is 10.2$\times10^{-8}~{\rm erg/cm}^{2}/{\rm s}$ derived from the peak flux of these 10 bursts.
  }
  \label{fig_burst_color_factor}
\end{figure}

{\bf
We  notice that the peak luminosities of the last two bursts in this work are close to the Eddington limit, which are much brighter than other double-peaked no-PRE bursts. Therefore, the double-peaked structure of the last two bursts in the LE light curves could be, to some extent, attributed to the high temperature of the burst emission around the burst peak time and the passband limitation of the telescope. As depicted in the right panels of Figure \ref{fig_outburst_spec}, the peak of the burst emission is shifted to $\sim$ 10 keV, which is beyond the LE's most effective energy. However, the  explanation above can not be applied to the double-peaked profile observed in the ME light curves.

Additionally, we observed a flux variation around the burst peak. Since the burst peak flux is close to the Eddington limit, a marginal photospheric radius expansion could occur which may contribute to this variation \citep{Penninx1989, 2010ApJ...712..964G}.
The sharp increase around the burst peaks in Figure \ref{fig_burst_color_factor} {\bf (the color faction as a function of the burst flux, as described below)} also implies that the burst peak fluxes are close to the Eddington limit \citep{Suleimanov2011}.
}
%The last two bursts exhibited a high peak flux, reaching the peak flux of the PRE bursts detected from this source, and displayed double-peaked lightcurves in the LE/ME lightcurves.
However,  there was no evidence of radius expansion or temperature decrease during bursts.
Their temperature and radius profiles showed a significant modulation, which could be caused by the modulation of the color factor {\bf \citep{2012ApJ...747...77G}}, i.e., $R_{\rm bb}=R_{\rm NS}/f_{\rm c}^{2}$, where $R_{\rm bb}$ is the apparent radius derived from the spectral fitting, $R_{\rm NS}$ is the effective radius and $f_{\rm c}$ is the color factor related to the scattering of the atmosphere. {\bf The color factor $f_{\rm c}$ as a function of the burst flux is plotted in  Figure \ref{fig_burst_color_factor}}, assuming the NS radius $R_{\rm NS}$=10 km and $L_{\rm Edd}$=10.2$\times10^{-8}~{\rm erg/cm}^{2}/{\rm s}$ (i.e., $L_{\rm peak}/L_{\rm Edd}$=1) derived from the peak flux of these 10 bursts,  $f_{\rm c}$=1.7 is estimated at the burst peak flux and $f_{\rm c}$=$1.7\times(R_{\rm tail}/R_{\rm peak})$=1.2 is estimated at the burst tail with $L_{\rm bb}/L_{\rm Edd}$=0.14. The vast difference between the two $f_{\rm c}$ values above is favored by a high abundance of heavy elements in the burning material \citep{Suleimanov2011}, i.e., $Z>Z_{\odot}=0.0134$.

{\bf
Alternatively,  the flux fluctuation near the Eddington limit could be related to a burst-induced corona \citep{1987ApJ...315L..43M}.
In this scenario, the burst emission evaporates the surface of the accretion disk, forming a corona, which then scatters the burst photons.
However, this model is valid only under specific conditions, such as a line of sight that barely grazes the edge of the disk, e.g., at narrow inclination angles of $75^{\circ}$ to $80^{\circ}$ or for a disk with a significant height.
Nevertheless, this model does not fully explain the observed hard X-ray deficit.
Another plausible explanation is that the material scattering the burst photons could be generated by the burst itself \citep{1987ApJ...315L..43M}.
}

To comprehensively analyze the characteristics of the non-PRE double-peaked bursts discussed here, we conducted a comparative study with a relevant set of similar bursts from sources such as 4U~1636--536 \citep{Bhattacharyya2006a,Zhang2009,Li2021a}, GX~17+2 \citep{Kuulkers2002},  XTE~J1709--267 \citep{Jonker2004},  MXB~1730--335 (Rapid Burster)\citep{Bagnoli2004}, GRS~1741.9--2853 \citep{Pike2021}, and other double-peaked no-PRE bursts from 4U~1608--52 \citep{Penninx1989,Jaisawal2019,Guver2021}.
We identified a similar burst profile reported by \citet{Penninx1989}, i.e., a bright burst with peak flux up to 1$\times10^{-7}~{\rm erg/cm}^{2}/{\rm s}$ and a several-seconds-long   $\sim$25\% dip of the bolometric flux near the burst peak. Possible explanations for this behavior include
{\bf the stall of thermonuclear flame spreading (e.g., ignition at high latitude but stalls on the equator) \citep{Bhattacharyya2006a,2006ApJ...641L..53B},  nuclear waiting impedance in rp-process \citep{2004ApJ...608L..61F}, or
 two-step generation/release of the thermonuclear energy caused by shear instabilities in the NS outer envelope \citep{1985ApJ...299..487S,1988A&A...199L...9F}. }
 %{\bf(e.g., \citealp{Guver2021})}.

\subsection{The saturation of the corona cooling by bursts: a hint of the geometry of the corona}

The hard X-ray drop has been attributed to the burst-induced coronal cooling.
This is supported by the decrease of the corona temperature from 60 keV to 20 keV observed in GS~1826--24 by stacking roughly 100 bursts from INTEGRAL observations \citep{Sanchez2020}.
In this scenario, it is expected that the deficit fraction would be higher in a higher energy band (e.g., \citealp{Chen1816}).
In our study, the trend is indeed observed at $<$ 50 keV both for the first 8 bursts and the last 2 bursts.
However, this trend ceases at $>$ 50 keV for the first 8 bursts, while it continues for the last 2 bursts.
The cooling saturation in the first 8 bursts may suggest the presence of another
mechanism or region responsible for hard X-ray generation that remains unaffected by the burst emission.

\begin{figure}
\centering
\includegraphics[angle=0, scale=0.4]{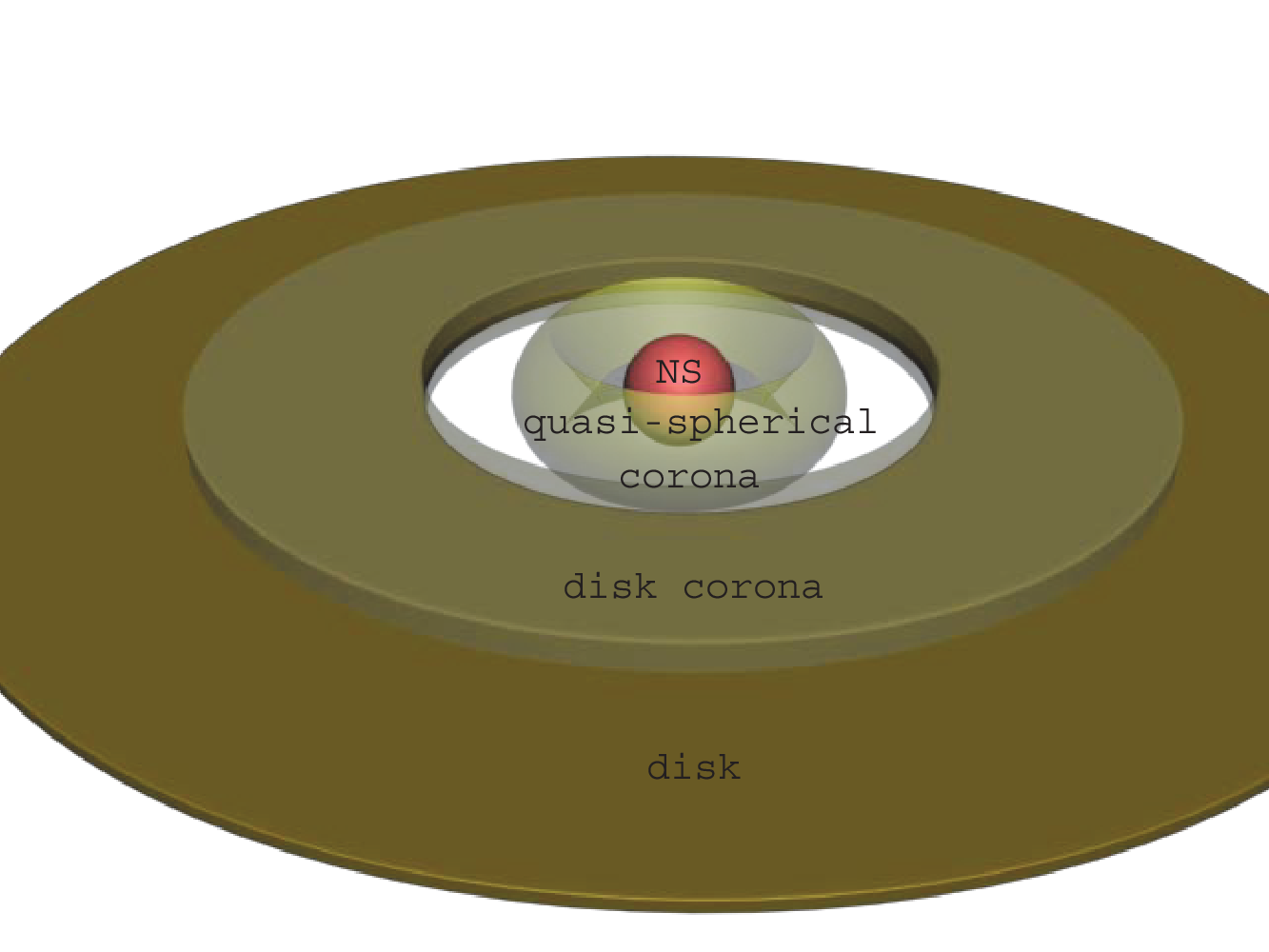}
  \caption{The illustration of the central region of a NS XRB, in which a disk corona is located around the disk and a quasi-spherical corona is around the NS.
  During a burst, the burst emission could escape from the NS surface without suffering severe up-scattering by the quasi-spherical corona. Meanwhile, at the burst peak, the quasi-spherical corona is cooled by the soft X-rays from the NS surface, but the disk corona is intact during the burst.
  }
\label{fig_illustration}
\end{figure}

We also notice that the characteristic pattern of the color factor $f_{\rm c}$  aligns with the atmosphere model prediction, with $f_{\rm c}$ decreasing in the early cooling phase.
This indicates that the burst emission  can escape from the NS surface without suffering severe Comptonization, suggesting the absence of the SL.
This supports the idea that the corona is a hot, optically thin quasi-spherical flow situated between the NS and the disk \citep{Kajava2014}, as illustrated in
Figure \ref{fig_illustration}.
The quasi-spherical corona, being near the NS surface, is initially cooled by the emergent burst emission, while another hard X-ray generation or region, such as the disk-corona, remains intact during bursts, leading to the observed hard X-ray saturation.

The structure of the two coronas is reminiscent of the warm-layer \citep{Zhang2000} and the disk-corona.
Another hard X-ray region, which remains intact from burst cooling, could be the jet, unaffected by the burst emission.
This two-corona scenario was also proposed by \citet{2005ApJ...634.1261T} with a spherical corona at $T_{\rm e}\sim$ 7 keV and a disk corona at $T_{\rm e}\sim$ 20 keV.
They suggested that the spherical corona's $T_{\rm e}$ was cooled by the burst to $\sim$ 3 keV, while the disk corona remained intact.

Furthermore, it appears that the hard X-ray emission fails to return to the pre-burst level, even during the final stages of the bursts.
{\bf In other words, the recovered timescale of the quasi-spheric corona is more than 80 s, consistent with the estimation from GS~1826--238 \citep{2005ApJ...634.1261T}.}
For the first 8 bursts, we extracted the spectrum of LE in 1.5--10 keV from 80 s to 85 s (with the burst peak time set as 0 s) and fitted it using the absorbed blackbody model.
 The derived bolometric burst flux is $\sim 1\times10^{-9}{\rm erg/cm}^{2}/{\rm s}$,
i.e., there is still a {\bf deficit} of the hard X-ray emission by a factor of roughly one quarter with the burst declining to 2\% of its peak flux.
In other words, assuming that the spectral shape of the persistent/pre-burst emission is not changed during bursts, the burst emission and the {\bf deficit} of the corona emission both is comparable, both are  $\sim 1\times10^{-9}{\rm erg/cm}^{2}/{\rm s}$.
The classical approach to X-ray burst spectroscopy, as used in this study, may lead to significantly overestimated burst temperatures if the spectral shape of the persistent/pre-burst emission remains unchanged during bursts. In such cases, parameters derived from the tail of the burst in low/hard states, such as the mass and radius of the NS \citep{2017AA...608A..31N}, must consider the Comptonization by the corona.

%Additionally, the recovered timescale of the quasi-spheric corona is more than 80 s, consistent with the estimation from GS~1826--238 \citep{2005ApJ...634.1261T}. For the spheric corona around NS, $T_{\rm e}$ takes $\sim$ 100--150 s and hundreds of seconds to recover to its one-half value and the whole value, respectively \citep{2005ApJ...634.1261T}.

%It is possible that the pre-burst/persistent spectra are well described by the disk-corona scenario and lack the quasi-spherical corona, which differs from the corona geometry derived from the burst hard X-ray {\bf deficit}. This discrepancy could be due to the data quality issues, e.g., lack of  photons below 1.5 keV and low count rate due to the small effective area of LE.

%Moreover, while bursts in the fainter lagging low/hard state of IGR~J17473--2721 did not show the hard X-ray {\bf deficit}, this was attributed to the low count rate of persistent/pre-burst emission in the hard X-ray band or a distant corona from the NS \citep{Chen2012}.  However, in this work, we observed the hard X-ray {\bf deficit} in bursts occurring in the fainter lagging low/hard state of 4U~1608--52.  This suggests that the absence of the hard X-ray {\bf deficit} of IGR~J17473--2721 could be due to the low count rate of the pre-burst/persistent emission in the hard X-ray band.

%\acknowledgements
\section{acknowledgements}
%We thank the reviewer for the constructive feedback and comments that greatly improved the quality of this paper.
This work made use of the data and software from the Insight-HXMT
mission, a project funded by China National Space Administration
(CNSA) and the Chinese Academy of Sciences (CAS).
This research has made use of data and software provided by of data obtained from the High Energy Astrophysics Science Archive Research Center (HEASARC), provided by NASA’s
Goddard Space Flight Center.
This work is supported by the National Key R\&D Program of China (2021YFA0718500) and the National Natural Science Foundation of China under grants 12173103, U2038101, 12233002.
This work was partially supported by International Partnership Program of Chinese Academy of Sciences (Grant No.113111KYSB20190020).

\bibliographystyle{plainnat}

\begin{thebibliography}{99}
%\bibitem[Arnaud (1996)]{Arnaud1996}Arnaud K. A., 1996, in Jacoby G. H., Barnes J., eds, Astronomical Society of the Pacific Conference Series Vol. 101, Astronomical Data Analysis Software and Systems V. p. 17
%\bibitem[Armas Padilla et al. (2017)]{ArmasPadilla2017}Armas Padilla, M., Ueda, Y., Hori, T., Shidatsu, M., Munoz-Darias, T., 2017, MNRAS, 467, 290
%\bibitem[Ballantyne \& Strohmayer (2004)]{Ball2004}Ballantyne, D. R., \& Strohmayer, T. E. 2004, ApJL, 602, L105
%\bibitem[Belian et al. (1976)]{Belian}Belian, R. D., Conner, J. P., \& Evans, W. D. 1976, ApJ, 206, L135
\bibitem[Bhattacharyya \& Strohmayer (2006a)]{Bhattacharyya2006a}Bhattacharyya, S. \& Strohmayer, T. E. 2006, ApJ, 636, 121
\bibitem[Bhattacharyya \& Strohmayer(2006b)]{2006ApJ...641L..53B} Bhattacharyya, S. \& Strohmayer, T.~E.\ 2006, ApJL, 641, L53. doi:10.1086/503768

\bibitem[Bagnoli et al. (2004)]{Bagnoli2004}Bagnoli, T., in’t Zand, J. J. M., Patruno, A., Watts, A. L., 2014, MNRAS, 437, 2790


%\bibitem[Bhattacharyya et al. (2018)]{Bhattacharyya2018}Bhattacharyya, S. Yadav, J. S.,  Sridhar, Navin, et al. 2018, ApJ, 860, 88
%\bibitem[Bult et al. (2021)]{Bult2021}Bult, P., Altamirano, D., Arzoumanian, Z. et al. 2021, ApJ, 920, 59




\bibitem[Chen et al. (2012)]{Chen2012} Chen, Y. P., Zhang, S., Zhang, S. N., et al. 2012, ApJL, 752, 34
\bibitem[Chen et al. (2013)]{Chen2013} Chen, Y. P., Zhang, S., Zhang, S. N., et al. 2013, ApJL, 777, 9
\bibitem[Chen et al. (2018)]{Chen2018} Chen, Y. P., Zhang, S., Qu, J. L., Zhang, S. N., et al. 2018, ApJL, 864, 30
\bibitem[Chen et al. (2019)]{Chen16081} Chen, Y. P., Zhang, S., Zhang, S. N., et al. 2019, Journal of High Energy Astrophysics, 24, 23
\bibitem[Chen et al. (2022)]{Chen1816} Chen, Y. P., Zhang, S., Ji, L., Zhang, S. N., et al. 2022, ApJL, 936, L12
\bibitem[Degenaar et al.(2016)]{2016MNRAS.456.4256D} Degenaar, N., Koljonen, K.~I.~I., Chakrabarty, D., et al.\ 2016, MNRAS, 456, 4256. doi:10.1093/mnras/stv2965

%\bibitem[Chen et al. (2022a)]{Chen2022a} Chen, Y. P., Zhang, S., Ji, L., Zhang, S. N., et al. 2022a, ApJ, 936, 46
%\bibitem[Chen et al. (2022b)]{Chen2022b} Chen, Y. P., Zhang, S., Ji, L., Zhang, S. N., et al. 2022b, arXiv:2209.10721
%\bibitem[Cominsky et al. (1978)]{Cominsky1978}Cominsky, L., Jones, C., Forman, W., \& Tananbaum, H. 1978, ApJ, 224, 46
%\bibitem[Cumming (2004)]{Cumming}Cumming, A. \ 2004, Nucl. Phys. B Proc. Suppl., 132, 435
%\bibitem[Degenaar et al.(2015)]{Degenaar2015}Degenaar, N., Miller, J. M., Chakrabarty, D., Harrison, F. A., Kara, E., \& Fabian, A. C. 2015, MNRAS, 451, L85
%\bibitem[Degenaar et al.(2018)]{Degenaar2018}Degenaar, N., Ballantyne, D. R., Belloni, T., et al. 2018, SSRv, 214, 15
\bibitem[Done et al. (2007)]{Done2007}Done, C., Gierliński, M., Kubota, A. 2007, A\&AR, 15, 1

\bibitem[Fisker et al.(2004)]{2004ApJ...608L..61F} Fisker, J.~L., Thielemann, F.-K., \& Wiescher, M.\ 2004, ApJL, 608, L61. doi:10.1086/422215
\bibitem[Fujimoto et al.(1988)]{1988A&A...199L...9F} Fujimoto, M.~Y., Sztajno, M., Lewin, W.~H.~G., et al.\ 1988, A\&A, 199, L9
%\bibitem[Forman et al. (1978)]{Forman1978}Forman, W., Jones, C., Cominsky, L., et al. 1978, ApJS, 38, 357
%\bibitem[Fragile et al.(2020)]{Fragile2020}Fragile, P. C., Ballantyne, D. R., \& Blankenship, A. 2020, NatAs, 4, 541
%\bibitem[G$\ddot{u}$ver et al.(2012)]{Guver2012}G$\ddot{u}$ver, T., Psaltis, D., \& Zel, F. 2012, ApJ, 747, 76
\bibitem[Galloway et al.(2008)]{Galloway}Galloway, D. K., Muno, M. P., Hartman, J. M., et al. \ 2008, ApJS, 179, 360
%\bibitem[Galloway, $\ddot{o}$zel \& Psaltis (2008a)]{Galloway2008a}Galloway, D. K., $\ddot{o}$zel, F., Psaltis, D. 2008a, MNRAS, 387, 268
\bibitem[Galloway et al. (2020)]{Galloway2020}Galloway, D. K., In’t Zand, J., Chenevez, J., et al. 2020, ApJS, 249, 32
\bibitem[Galloway \& Keek (2021)]{Galloway2021}Galloway, D. K., \& Keek, L. 2021, Astrophys. Space Sci. Lib., 461, 209
\bibitem[Gilfanov et al. (2003)]{Gilfanov2003}Gilfanov, M.; Revnivtsev, M.; Molkov, S.Gilfanov et al. 2003, A\&A, 410, 217
\bibitem[Grebenev \& Sunyaev (2002)]{Grebenev2002}Grebenev, S. A., \& Sunyaev, R. A. 2002, AstL, 28, 150

\bibitem[G{\"u}ver et al.(2022)]{2022ApJ...935..154G} G{\"u}ver, T., Bostanc{\i}, Z.~F., Boztepe, T., et al.\ 2022, \apj, 935, 154. doi:10.3847/1538-4357/ac8106
\bibitem[Güver et al. (2021)]{Guver2021}Güver, T., Boztepe, T., Göğüş, E. et al. 2021, ApJ, 910, 37
\bibitem[G{\"u}ver et al.(2012)]{2012ApJ...747...77G} G{\"u}ver, T., {\"O}zel, F., \& Psaltis, D.\ 2012, \apj, 747, 77. doi:10.1088/0004-637X/747/1/77

\bibitem[G{\"u}ver et al.(2010)]{2010ApJ...712..964G} G{\"u}ver, T., {\"O}zel, F., Cabrera-Lavers, A., et al.\ 2010, \apj, 712, 964. doi:10.1088/0004-637X/712/2/964


%\bibitem[in't Zand et al. (1999)]{int1999}in't Zand, J. J. M., Heise, J., Kuulkers, E., et al. 1999, A\&A, 347, 891
%\bibitem[He \& Keek (2016)]{He2006}He, C.-C. \& Keek, L. 2016, ApJ, 819, 47
%\bibitem[in't Zand et al. (2013)]{int2013}in't Zand, J. J. M., Galloway, D. K., Marshall, H. L., et al. 2013, A\&A,553, A83
\bibitem[Jaisawal et al. (2019)]{Jaisawal2019}Jaisawal, G. K., Chenevez, J., Bult, P., et al. 2019, ApJ, 883, 61


\bibitem[Ji et al. (2013)]{Ji2013}Ji, L., Zhang, S., Chen, Y. P., et al. 2013, MNRAS, 432, 2773
\bibitem[Ji et al. (2014a)]{Ji2014a}Ji, L., Zhang, S., Chen, Y. P., et al., 2014, ApJ, 791, L39
\bibitem[Ji et al. (2014b)]{Ji2014b}Ji, L., Zhang, S., Chen, Y. P., et al., 2014, A\&A, 564, A20
\bibitem[Kajava et al.(2017)]{2017AA...599A..89K} Kajava, J.~J.~E., S{\'a}nchez-Fern{\'a}ndez, C., Kuulkers, E., et al.\ 2017, A\&A, 599, A89. doi:10.1051/0004-6361/201629542


\bibitem[Jonker et al. (2004)]{Jonker2004}Jonker P. G., Galloway D. K., McClintock J. E., Buxton M., Garcia M., Murray S., 2004, MNRAS, 354, 666

%\bibitem[Kaastra \& Bleeker (2016)]{Kaastra2016}Kaastra, J. S.; Bleeker, J. A. M. 2016, A\&A, 587, A151
%\bibitem[Kajava et al. (2014)]{Kajava2014}Kajava, J. J. E., Nättilä, J., Latvala, O. M., et al. 2014, MNRAS, 445, 4218
\bibitem[Kajava et al.(2014)]{Kajava2014} Kajava, J.~J.~E., N{\"a}ttil{\"a}, J., Latvala, O.-M., et al.\ 2014, MNRAS, 445, 4218. doi:10.1093/mnras/stu2073

\bibitem[Kashyap et al. (2022)]{Kashyap2022}Kashyap, U., Ram, B., Guver, T., Chakraborty, M. 2022, MNRAS, 509, 3989
%\bibitem[Kajava et al. (2017)]{Kajava2017}Kajava, J. J. E., Koljonen, K. I. I., N$\ddot{a}$ttil$\ddot{a}$, J., Suleimanov, V., \& Poutanen, J. 2017, MNRAS, 472, 78
%\bibitem[Keek et al. (2014)]{Keek2014}Keek, L., Ballantyne, D. R., Kuulkers, E., \& Strohmayer, T. E. 2014, ApJL, 797, L23
%\bibitem[Keek et al. (2014)]{KeeK2014b}Keek, L., Ballantyne, D. R., Kuulkers, E., \& Strohmayer, T. E. 2014b, ApJL, 797, L23
%\bibitem[Keek et al. (2018)]{Keek2018}Keek, L., Arzoumanian, Z., Bult, P., et al. 2018, ApJL, 855,4
%\bibitem[Keek et al. (2018a)]{Keek2018a}Keek, L., Arzoumanian, Z., Chakrabarty, D., et al. 2018, ApJL, 856,37
\bibitem[Kuulkers et al. (2002)]{Kuulkers2002}Kuulkers, E., Homan, J., van der Klis, M et al., 2002, A\&A, 382, 947
%\bibitem[Kuulkers et al. (2003)]{Kuulkers2003}Kuulkers, E., den Hartog, P. R., in ’t Zand, J. J. M., et al. 2003, A\&A 399, 663
%\bibitem[Lewin et al.(1993)]{Lewin}Lewin, W. H. G., van Paradijs, J., \& Taam, R. E. \ 1993, Space Sci. Rev., 62, 223
%\bibitem[Li et al (2022)]{Li2022}Li, Z. S., Yu, W. H., Lu, Y. Q., et al. 2022, arXiv:2205.12037

\bibitem[Li et al. (2021)]{Li2021a}Li, C., Zhang, G., Méndez, M., Wang, J., \& Lyu, M. 2021, MNRAS, 501, 168


%\bibitem[Li et al. (2020)]{Li2020}Li, X. B., Li, X. F., Tan, Y. et al. 2020, JHEA, 27, 64
%\bibitem[Lin et al. (2009)]{Lin2009}Lin, D., Altamirano, D., Homan, J. et al. 2009,  ApJ, 699, 60
\bibitem[Maccarone \& Coppi (2003)]{maccarone2003}Maccarone, T. J. \& Coppi, P. S. \ 2003, A\&A, 399, 1151
\bibitem[Melia (1987)]{1987ApJ...315L..43M} Melia, F.\ 1987, ApJL, 315, L43. doi:10.1086/184858

%\bibitem[Muno et al. (2001)]{Muno}Muno, M. P., Chakrabarty, D., Galloway, D. K., \& Savov, P. 2001, ApJ, 553, L157
\bibitem[N{\"a}ttil{\"a} et al.(2017)]{2017AA...608A..31N} N{\"a}ttil{\"a}, J., Miller, M.~C., Steiner, A.~W., et al.\ 2017, ApJ, 608, A31. doi:10.1051/0004-6361/201731082

\bibitem[Penninx et al. (1989)]{Penninx1989}Penninx W., Damen, E., Tan, J. et al. 1989, A\&A, 208, 146



%\bibitem[PenninxW et al. (1989)]{PenninxW1989}PenninxW., Damen E., van Paradijs J., Tan J., Lewin W. H. G., 1989, A\&A, 208, 146
\bibitem[Pike et al. (2021)]{Pike2021}Pike, S. N., Harrison, F. A., Tomsick, J. A. et al. 2021, ApJ, 918, 9
%\bibitem[Poutanen et al. (2014)]{Poutanen}Poutanen, J., N$\ddot{ai}$ttil$\ddot{a}$, J., Kajava, J. J. E. et al. 2014, MNRAS, 442, 3777
%\bibitem[Remillard et al. (2022)]{Remillard2022}Remillard, R. A., Loewenstein, M., Steiner, J. F. et al. 2022, AJ, 163, 130
\bibitem[S{\'a}nchez-Fern{\'a}ndez et al. (2020)]{Sanchez2020}S{\'a}nchez-Fern{\'a}ndez C., Kajava J. J. E., Poutanen J., et al., 2020, A\&A, 634, A58
\bibitem[Speicher et al.(2020)]{2020MNRAS.499.4479S} Speicher, J., Ballantyne, D.~R., \& Malzac, J.\ 2020, MNRAS, 499, 4479
%\bibitem[Rybicki \& Lightman (2004)]{Rybicki2004}Rybicki, G. B., \& Lightman, A. P. 2004, Radiative Processes in Astrophysics (Weinheim: Wiley-VCH)
%\bibitem[Shaposhnikov \& Titarchuk (2004)]{Shaposhnikov2004}Shaposhnikov, N., \& Titarchuk, L. 2004, ApJ, 606, L57
%\bibitem[Shaposhnikov et al. (2003)]{Shaposhnikov2003}Shaposhnikov, N., Titarchuk, L., Haberl, F., 2003, ApJL, 593, L35
%\bibitem[Stahl et al. (2013)]{Stahl}Stahl, A., Klu$\acute{z}$niak, W., Wielgus, M., \& Abramowicz, M. 2013, A\&A, 555, A114
%\bibitem[Strohmayer \& Bildsten(2006)]{Strohmayer} Strohmayer, T., \& Bildsten, L.   \ 2006, New views of thermonuclear bursts (Compact stellar X-ray sources), 113, 156
\bibitem[Suleimanov \& Poutanen (2006)]{Suleimanov2006}Suleimanov, V., \& Poutanen, J. 2006, MNRAS, 369, 2036
\bibitem[Suleimanov et al. (2011)]{Suleimanov2011}Suleimanov, V., Poutanen, J., Werner, K. 2011, A\&A, 527, A139

\bibitem[Sztajno et al.(1985)]{1985ApJ...299..487S} Sztajno, M., van Paradijs, J., Lewin, W.~H.~G., et al.\ 1985, ApJ, 299, 487. doi:10.1086/163715

%\bibitem[Suleimanov et al. (2012)]{Suleimanov2012}Suleimanov, V., Poutanen, J., Werner, K. 2012, A\&A, 545, A120
%\bibitem[Suleimanov et al. (2018)]{Suleimanov2018}Suleimanov, V., Poutanen, J., Werner, K. 2018, A\&A, 619, A114
% \bibitem[Titarchuk \& Shaposhnikov  (2002)]{Titarchuk2002}Titarchuk, L. \& Shaposhnikov, N.  2002, ApJL, 570, L25
 %\bibitem[Titarchuk    (1994)]{Titarchuk1994}Titarchuk, L. 1994, ApJ, 429, 340
%\bibitem[Tomsick et al. (2007)]{Tomsick2007}Tomsick, J. A., Gelino, D. M., \& Kaaret, P. 2007, ApJ, 663, 461
\bibitem[Thompson et al.(2005)]{2005ApJ...634.1261T} Thompson, T.~W.~J., Rothschild, R.~E., Tomsick, J.~A., et al.\ 2005, ApJ, 634, 1261. doi:10.1086/497104

%\bibitem[Wilms et al. (2000)]{Wilms2000}Wilms, J., Allen, A., \& McCray, R. 2000, ApJ, 542, 914
%\bibitem[Worpel et al. (2013)]{Worpel2013}Worpel, H., Galloway, D. K., \& Price, D. J. 2013, ApJ, 772, 94
%\bibitem[Worpel et al. (2015)]{Worpel2015}Worpel, H., Galloway, D. K., \& Price, D. J. 2015, ApJ, 801, 60
%\bibitem[Zdziarski et al. (2020)]{Zdziarski2020}Zdziarski, A. A., Szanecki, M., Poutanen, J., Gierlinski, M., \& Biernacki, P. 2020, MNRAS, 492, 5234
\bibitem[Zhang et al. (2000)]{Zhang2000}Zhang, S. N., et al., 2000, Science, 287, 1239
\bibitem[Zhang et al. (2009)]{Zhang2009}Zhang, G. B., M$\acute{e}$ndez, M., Altamirano, D., Belloni, T. M., Homan, J., 2009, MNRAS, 398, 368
%\bibitem[Zhang et al. (2014)]{Zhang2014} Zhang, S.,  L,u F. J., Zhang, S. N. et al. in Space Telescopes and Instrumentation 2014: Ultraviolet to Gamma Ray, Proc. SPIE, Vol. 9144 (2014) p. 914421
\bibitem[Zhang et al. (2019)]{Zhang2019} Zhang, S. N., Santangelo, A., Feroci, M., et al.  2019, Science China Physics, Mechanics \& Astronomy, Volume 62, Issue 2, article id. 29502, 25
\bibitem[Zhang et al. (2020)]{Zhang2020}Zhang, S.-N., Li, T.-P., Lu, F.-J., et al. 2020, SCPMA, 63, 249502


\end{thebibliography}

\begin{landscape}

\begin{table}
\tiny
\begin{center}
\caption{Bursters with hard X-ray deficit}
\label{tb_burster}
 \begin{tabular}{lccccccll}
\\\hline
Source & $n_{\rm}$& Energy band & Significance & $F_{\rm per}$ &  $F_{\rm b}$ & Lag  & Instrument & Reference   \\
   & &  keV & $\sigma$ & cts/s & cts/s & s & \\\hline
Aql~X-1 & 1 & 30--60 & 2 & 10  & 5   & & RXTE/HEXTE & \citet{maccarone2003} \\
          & 21 & 40--50 & 6 & 0.4  & -0.5  & 1.8$\pm$ 1.5 & RXTE/PCA & \citet{Chen2013} \\\hline
IGR~17473-2721 & 40 &30--50 & 6 & 2 & 1 & 0.7$\pm$0.5  & RXTE/PCA & \citet{Chen2012}\\\hline
4U~1636-536 & 114&30--50 & 3 & 1.5 & 0.6 & 2.5$\pm$1.5  & RXTE/PCA & \citet{Ji2013}\\
            &1&40--70 & 6 & 15 & -1 & 1.6$\pm$1.2  & Insight-HXMT/HE & \citet{Chen2018}\\
            &7&30--79 & - &  $\sim$1 & $\sim$0.5 &    & NuSTAR & \citet{2022ApJ...935..154G}\\\hline
GS~1826–238 &43&30--50 & 21 & 1.7 & 0.8 & 3.6$\pm$1.2  & RXTE/PCA & \citet{Ji2014a}\\
          &90&35--70 &  - & 10 & 0 & -  & INTEGRAL/ISGRI & \citet{Sanchez2020}\\\hline
KS~1731-260 &16&40--50 & 4.5 & 0.3 & -0.6 & 0.9$\pm$2.1  & RXTE/PCA & \citet{Ji2014b}\\\hline
4U~1705-44 &30&40--50 & 4.7 & 0.3 & -0.3 & 2.5$\pm$2.0  & RXTE/PCA & \citet{Ji2014b}\\\hline
4U~1728-34 &123 &40--50 & 3.4 & $\sim$3 & $\sim$0 & - & INTEGRAL/ISGRI & \citet{2017AA...599A..89K} \\\hline
4U~1724-30 & 1 &30--80 & - &100 & 80 & -& AstroSat/LAXPC & \citet{Kashyap2022} \\\hline
MAXI~1816-195 & 66 & 30-100 & 15.7 & 40 & 28 & $\sim$ 1 & Insight-HXMT/HE & \citet{Chen1816}\\\hline
4U~1608--52 & 8 & 30-100 & 22.1 & 59 & 29 & $\sim$ 1 & Insight-HXMT/HE & Chen et al. (2024)\\\hline
   \end{tabular}
\begin{list}{}{}
\item[Note:]{The columns denote the source/burster name, number of bursts $n_{\rm}$ used for the derived the hard X-ray {\bf deficit}, the X-ray energy band used for the hard X-ray deficit, the significance of the deficit, the persistent/pre-burst count rate ($F_{\rm per}$), the count rate during burst ($F_{\rm b}$), the time lag of the hard X-ray deficit, the telescope used for the deficit detection and the reference.}
  \end{list}
  \end{center}
\end{table}
\end{landscape}

\begin{table}
\begin{center}
\caption{The bursts' parameters  of 4U~1608--52  detected by Insight/HXMT in the low/hard state of 2022 outburst: obsid, peak time $t_{\rm p}$, fluence   $E_{\rm b}$,  peak flux $F_{\rm p}$   derived from the blackbody model, burst duration $\tau$ from  $E_{\rm b}$/$F_{\rm p}$.
}
\label{tb_burst}
 %\vspace{5pt}
\begin{tabular}{cccccccccc}
\\\hline
  No & obsid & $t_{\rm p}$  & $E_{\rm b}$ & $F_{\rm p}$  & $\tau$   \\\hline
  &    	&  MJD   & a  &  b   & s    \\\hline
  1$^{*}$ & P050422200102-20220907-01-01 & 59829.233537 & $99.3_{-2.3}^{+2.3}$ & $5.22_{-0.48}^{+0.51}$ & $19.0_{-1.8}^{+1.9}$ \\\hline
2 & P050422200103-20220907-01-01 & 59829.447900 & $134.0_{-1.1}^{+1.1}$ & $6.73_{-0.29}^{+0.29}$ & $19.9_{-0.9}^{+0.9}$ \\\hline
3 & P050422200104-20220907-01-01 & 59829.658991 & $138.2_{-1.1}^{+1.1}$ & $6.19_{-0.28}^{+0.28}$ & $22.3_{-1.0}^{+1.0}$ \\\hline
4$^{*}$ & P050422200201-20220908-01-01 & 59830.685675 & $117.8_{-2.5}^{+2.5}$ & $6.45_{-0.53}^{+0.55}$ & $18.3_{-1.5}^{+1.6}$ \\\hline
5 & P050422200502-20220911-01-01 & 59833.765484 & $137.2_{-1.1}^{+1.1}$ & $6.32_{-0.28}^{+0.28}$ & $21.7_{-1.0}^{+1.0}$ \\\hline
6 & P050422200503-20220911-01-01 & 59833.957544 & $148.1_{-1.2}^{+1.2}$ & $6.24_{-0.29}^{+0.29}$ & $23.7_{-1.1}^{+1.1}$ \\\hline
7 & P050422200603-20220913-01-01 & 59835.853282 & $117.3_{-3.8}^{+4.1}$ & $5.58_{-0.29}^{+0.29}$ & $21.0_{-1.3}^{+1.3}$ \\\hline
8 & P050422200703-20220916-02-01 & 59838.251922 & $145.7_{-1.1}^{+1.1}$ & $6.14_{-0.28}^{+0.28}$ & $23.7_{-1.1}^{+1.1}$ \\\hline
9 & P050422200902-20220919-01-01 & 59841.622706 & $150.2_{-1.1}^{+1.1}$ & $9.80_{-0.35}^{+0.35}$ & $15.3_{-0.6}^{+0.6}$ \\\hline
10 & P050422201102-20220922-01-01 & 59844.361977 & $150.1_{-1.1}^{+1.1}$ & $10.19_{-0.35}^{+0.35}$ & $14.7_{-0.5}^{+0.5}$ \\\hline
 \end{tabular}
\end{center}
\begin{list}{}{}
\item[a:]{In units of $10^{-8}~{\rm erg}~{\rm cm}^{2}$}
\item[b:]{In units of $10^{-8}~{\rm erg}~{\rm cm}^{2}~{\rm s}^{-1}$}
\item[$^{*}$:]{The bursts absented from LE data}
 \end{list}
\end{table}

\begin{table}%\tiny
\centering
\caption{The results of  spectral fit of the   persistent emission of burst \#1--\#8  with  cons*tbabs*thcomp*diskbb }
\label{tb_fit_thcomp_diskbb}
%\vskip -0.4cm
\begin{tabular}{cccccccccccc}
\\\hline
No &Time   & $\tau$ & $kT_{\rm e}$  & $f_{\rm sc}$ & $kT_{\rm disk}$ & $R_{\rm disk}$ & $F_{\rm disk}$ &$F_{\rm corona}$ &$F_{\rm total}$ & $\chi_{\nu}^{2}$ (d.o.f.)\\
   & MJD  &  & keV & &keV & km & $10^{-9}$  & $10^{-9}$  & $10^{-9}$ & \\
\hline
 1-4  & 59829.06   & $3.3_{-0.2}^{+0.3}$ & $25.18_{-2.62}^{+1.54}$ & $1.00_{-0.2}^{}$ & $0.43_{-0.09}^{+0.10}$ & $17.1_{-4.5}^{+8.5}$ & $0.99_{-0.00}^{+0.00}$ & $4.36_{-0.03}^{+0.03}$ & $5.35_{-0.03}^{+0.03}$ & $1.00(60)$ \\\hline
5-6  & 59833.57   & $2.7_{-0.3}^{+0.2}$ & $30.74_{-4.47}^{+4.96}$ & $1.00_{-0.0}^{}$ & $0.48_{-0.06}^{+0.07}$ & $14.9_{-3.9}^{+10.5}$ & $1.15_{-0.01}^{+0.01}$ & $4.39_{-0.03}^{+0.03}$ & $5.54_{-0.03}^{+0.03}$ & $1.11(60)$ \\\hline
7  & 59835.55   & $3.2_{-0.3}^{+0.5}$ & $24.20_{-4.16}^{+4.80}$ & $0.98_{-0.1}^{}$ & $0.50_{-0.08}^{+0.10}$ & $13.5_{-3.2}^{+7.0}$ & $1.13_{-0.01}^{+0.01}$ & $3.98_{-0.03}^{+0.03}$ & $5.11_{-0.03}^{+0.03}$ & $1.08(57)$ \\\hline
8  & 59838.07   & $2.0_{-0.4}^{+0.3}$ & $47.26_{-9.01}^{+9.92}$ & $1.00_{-0.0}^{}$ & $0.52_{-0.10}^{+0.08}$ & $11.9_{-2.7}^{+4.6}$ & $0.99_{-0.01}^{+0.01}$ & $3.49_{-0.03}^{+0.03}$ & $4.48_{-0.03}^{+0.02}$ & $1.11(59)$ \\\hline
 9 & 59841.71   & $3.2_{-0.9}^{+0.7}$ & $25.64_{-5.91}^{+13.18}$ & $0.76_{-0.1}^{+0.1}$ & $0.65_{-0.08}^{+0.09}$ & $7.4_{-1.4}^{+2.5}$ & $0.94_{-0.01}^{+0.01}$ & $2.55_{-0.03}^{+0.03}$ & $3.49_{-0.03}^{+0.03}$ & $0.57(56)$ \\\hline
 10 & 59844.38   & $2.6_{-0.7}^{+0.5}$ & $34.63_{-7.41}^{+16.53}$ & $0.72_{-0.1}^{+0.1}$ & $0.69_{-0.06}^{+0.06}$ & $6.8_{-0.9}^{+1.4}$ & $1.00_{-0.01}^{+0.01}$ & $2.61_{-0.03}^{+0.03}$ & $3.61_{-0.03}^{+0.03}$ & $1.37(58)$ \\\hline
\end{tabular}
\begin{list}{}{}
\item[a]{: The model parameters: the optical depth $\tau$,  the  electron temperature $kT_{\rm e}$,  the cover factor $f_{\rm sc}$,  the accretion disk  temperature   $kT_{\rm disk}$ and  the inner disk radius   $R_{\rm diskb}$ at a distance of 4 kpc and an inclination angel $\theta$=30$^{\circ}$, the bolometric flux of the diskbb $F_{\rm diskbb}$, the bolometric flux of the corona $F_{\rm corona}$, and the total  bolometric flux  $F_{\rm total}$ are in units of $10^{-9}~{\rm erg/cm}^{2}/{\rm s}$, reduced $\chi_{\nu}^{2}$ and the degree of freedom (d.o.f.).}
 \end{list}
\end{table}

\clearpage

% \begin{figure}[t]
 %  \includegraphics[angle=270, scale=0.5]{lc_nicer_le_me.eps}
% \caption{
% Lightcurves of   4U~1608--52 with time bin 1 s. The top, middle and bottom panel is NICER, LE and ME   in their full energy band respectively. No background is subtracted.  }
%\label{fig_lc_nicer_le_me}
%\end{figure}

%\begin{figure}[t]
%\centering
%\includegraphics[angle=0, scale=0.3]{P040420700504-20220720-01-01_fit_bb_fit_flux.eps}
%\includegraphics[angle=0, scale=0.3]{P040420701503-20220809-01-01_fit_bb_fit_flux.eps}
% \caption{Fitting results of the temporal evolution of the bolometric flux of burst \#2 (left) and burst \#9 (right) with an exponential decay function (red) and a   power-law function (blue).  }
%\label{fig_burst_fit_bb_fit_flux}
%\end{figure}

%4U 1608 8 bursts in 2022 low/hard state
%root [3] 128828/3.868833E+02
%(const double)3.32989301941955091e+02
%root [4] 136290/3.869229E+02
%(const double)3.52240717724383842e+02
%root [5] (136290-128828)/sqrt(136290+128828)
%(const double)1.44922439938467598e+01    //significance

%\begin{figure}[t]
%\centering
%\includegraphics[angle=-90, scale=0.5]{xspec_le_me_he_fixed_2.eps}
%  \caption{The stacked burst spectrum fitted with model cons*wabs*(bbodryad+bbodyrad). The energy ranges of the spectrum of LE (black), ME (red) and HE (green) are 1--10 keV, 8–30 keV and 30–100 keV, respectively. Please note that only the LE data and the ME data are used in the spectral fitting, the HE data are not involved because of their negative values; i.e., the HE data are just for display purposes--indicating the hard X-ray {\bf deficit}. %  since the negative value cannot been shown in the main panel with a logarithmic coordinate axis  }
%\label{fig_burst_spec_le_me_he}
%\end{figure}

\label{lastpage}
\end{document}